\documentclass[aps,pre,superscriptaddress,twocolumn,tightenlines,showpacs,amsmath,amssymb,usenames,dvipsnames]{revtex4-1}

\usepackage[utf8]{inputenc} 
\usepackage[T1]{fontenc}

\usepackage{array}
\usepackage{amsmath}
\usepackage{amssymb}
\usepackage{bm}
\usepackage{bbm}
\usepackage{float}
\usepackage{tikz} 
\usepackage{graphicx}
\usepackage{booktabs}
\usepackage{dcolumn}
\usepackage[normalem]{ulem}
\usepackage{float}
\usepackage{mathtools}
\usepackage{hyperref}
\hypersetup{
    colorlinks,
    linkcolor=blue,
    citecolor=blue,
    filecolor=blue,
    urlcolor=blue
}
\usepackage{siunitx}
\usepackage{mathdots}
\usepackage{stmaryrd}
\usepackage{tikz}
\usepackage{tikz-cd}
\usepackage{gnuplot-lua-tikz}
\usepackage{mathtools}
\ifcsname gpsetdashtype\endcsname
\else
\newcommand{\gpsetdashtype}[1]{} 
\fi

\usepackage[outdir=./]{epstopdf}

\usepackage[caption=false]{subfig}

\newcommand{\mean}[1]{\langle #1\rangle}
\newcommand{\mb}[1]{\mathbf{#1}}

\newcommand{\bds}[1]{\boldsymbol{#1}}

\newcommand{\abs}[1]{\vert #1\vert}

\newcommand{\zz}{\textcolor{lightgray}{0}}

\DeclareMathOperator*{\Tr}{Tr}

\makeatletter
\newcolumntype{B}[3]{>{\boldmath\DC@{#1}{#2}{#3}}c<{\DC@end}}
\makeatother

\newcolumntype{C}{>{$}c<{$}}
\AtBeginDocument{
\heavyrulewidth=.08em
\lightrulewidth=.05em
\cmidrulewidth=.03em
\belowrulesep=.65ex
\belowbottomsep=0pt
\aboverulesep=.4ex
\abovetopsep=0pt
\cmidrulesep=\doublerulesep
\cmidrulekern=.5em
\defaultaddspace=.5em
}

\begin{document}

\title{Identification of emergent constraints and hidden order in frustrated magnets using tensorial kernel methods of machine learning}

\author{Jonas Greitemann}
\author{Ke Liu}
\email{ke.liu@lmu.de}
\affiliation{Arnold Sommerfeld Center for Theoretical Physics, University of Munich, Theresienstr. 37, 80333 München, Germany}
\affiliation{Munich Center for Quantum Science and Technology (MCQST), Schellingstr. 4, 80799 München, Germany}

\author{Ludovic D.C. Jaubert}
\affiliation{CNRS, Université de Bordeaux, LOMA, UMR 5798, 33400 Talence, France}

\author{Han Yan}
\author{Nic Shannon}
\affiliation{Theory of Quantum Matter Unit, Okinawa Institute of Science and Technology Graduate University, Onna-son, Okinawa 904-0412, Japan}

\author{Lode Pollet}
\affiliation{Arnold Sommerfeld Center for Theoretical Physics, University of Munich, Theresienstr. 37, 80333 München, Germany}
\affiliation{Munich Center for Quantum Science and Technology (MCQST), Schellingstr. 4, 80799 München, Germany}
\affiliation{Wilczek Quantum Center, School of Physics and Astronomy, Shanghai Jiao Tong University, Shanghai 200240, China}

\date{\today}

\begin{abstract}
Machine-learning techniques have proved successful in identifying ordered phases of matter.
However, it remains an open question how far they can contribute to the understanding of phases without broken symmetry, such as spin liquids.
Here we demonstrate how a machine learning approach can automatically learn the intricate phase diagram of a classical frustrated spin model.
The method we employ is a support vector machine equipped with a tensorial kernel and a spectral graph analysis which admits its applicability in an effectively unsupervised context.
Thanks to the interpretability of the machine we are able to infer, in closed form, both order parameter tensors of phases with broken symmetry, and the local constraints which signal an emergent gauge structure, and so characterize classical spin liquids.
The method is applied to the classical XXZ model on the pyrochlore lattice where it distinguishes---among others---between a hidden biaxial spin nematic phase and several different classical spin liquids.
The results are in full agreement with a previous analysis by Taillefumier \emph{et al.}  [Phys. Rev. X 7,  041057 (2017)], but go further by providing a systematic hierarchy between disordered regimes, and establishing the physical relevance of the susceptibilities associated with the local constraints.
Our work paves the way for the search of new orders and spin liquids in generic frustrated magnets.
\end{abstract}

\maketitle

\section{Introduction}
The theoretical study of frustrated models has a long and distinguished history~\cite{Bernal33,Pauling35,Wannier50,Anderson56}.
Nonetheless, the range of different phenomena seen in experiments on frustrated magnets still greatly outnumbers the number of problems solved in theory.
And, while numerical simulation techniques have become very sophisticated, in the presence of a spin liquid~\cite{Balents10, Savary16b,ZhouNg17} or spin nematic~\cite{Andreev84,Chubukov90,Chalker92,MomoiShannon06}, the most that is likely to be revealed by a conventional analysis 
is that the system lacks any conventional order.
For this reason, the phase diagram of each new model of a frustrated magnet must still be pieced together ``by hand'', a process which depends heavily on the ingenuity of individual researchers, and commonly takes decades.

State-of-the-art synthesis and characterization of new materials takes months, not years, whereas the time it takes to solve frustrated models is a serious bottleneck.
And it is therefore interesting to ask whether the techniques of machine learning, which have shown great promise when applied to different forms of magnetic order~\cite{CarrasquillaMelko17,PonteMelko17,Greitemann19,Liu19}, can also be used to construct the phase diagram of a highly frustrated magnet?

This question takes on conceptual, as well as practical, interest in the case of spin liquids.
Attempts to identify spin liquids by negation (``the system does not order'') lead quickly to a dead end, especially where more than one unconventional phase is at hand.
However, in the absence of a broken symmetries or thermodynamic phase transitions, researchers have traditionally struggled to find a satisfactory formulation of what distinguishes one state of matter from another.
How then would a machine, unencumbered by the weight of history and semantics, interpret its first spin liquid?
And how would it distinguish this spin liquid from a second, or a third, or from a different unconventional phase, with hidden order?

Two things are needed to answer these questions.
The first is a model, accessible to simulation, which is known to host a wide range of different phases, including some which are spin liquids and/or host unconventional order.
And the second is an \emph{interpretable} method of machine learning, so that once simulation results have been sorted into different phases, it is possible to interrogate the machine about the principle it used to classify each phase.

The first of these requirements is well met by the classical limit of the XXZ model on a pyrochlore lattice, which is accessible to large-scale classical Monte Carlo simulation, and known to support a plethora of different spin liquids, one of which also possesses hidden spin-nematic order~\cite{Taillefumier17,Benton18}.

To meet the second requirement, we turn to support vector machines (SVMs)~\cite{CortesVapnik95}, equipped with a tensorial kernel~\cite{Greitemann19,Liu19}.
These have already proved successful in identifying unconventional, hidden~\cite{Greitemann19} orders, and are fully interpretable, providing explicit forms for the order parameter in each case.
Furthermore, while SVMs technically fall under the umbrella of the supervised machine learning paradigm, we can still construct a graph encoding the similarity of the physics between any two points in the parameter space~\cite{Liu19}.
By subjecting it to a spectral graph analysis, we can infer the phase diagram in lieu of prior physical insight, effectively rendering the method \emph{semantically} unsupervised.
It is therefore a good starting point for exploring the still-more complex, and seemingly ambiguous, physics of spin liquids.

With these facts in mind, in this paper, we develop at length the application of the tensorial kernel SVM (TK-SVM) to the (classical) XXZ model on a pyrochlore lattice.
The results are encouraging.
Not only does the machine fully reproduce the complex phase diagram found by Taillefumier \emph{et al.}~\cite{Taillefumier17}, including finite-temperature crossovers between different forms of spin liquid, it also correctly identifies both the order parameters of broken-symmetry phases, and the \emph{constraints} on local spin configurations which characterize each different spin liquid.

\begin{figure}
  \centering
  \includegraphics[width=0.45\textwidth]{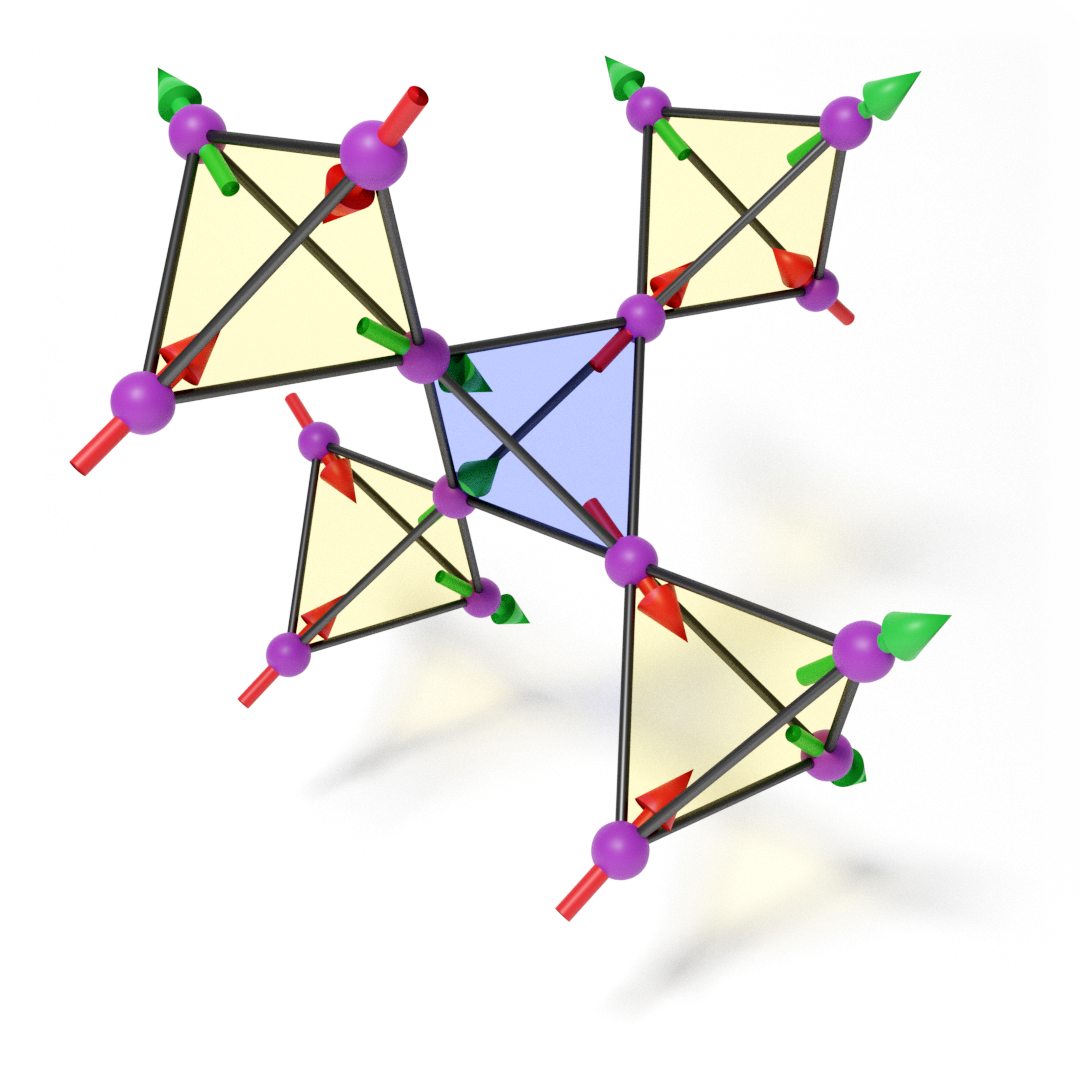}
  \vskip-0.8cm
  \caption{
  A spin configuration obeying the ``ice rules'', on a pyrochlore lattice built of corner-sharing tetrahedra.
  Such states do not have any long-range magnetic order, but satisfy a local constraint in which two spins point into, and two spins point out of, each tetrahedron \cite{Bernal33,Pauling35}.
  The spins in this rendering are aligned to their [111] direction (local $z$-axis), like the Ising spins in the spin-ice materials $\mathrm{Ho_2Ti_2O_7}$~\cite{Harris97}
  and $\mathrm{Dy_2Ti_2O_7}$~\cite{Bramwell01}.
  }
  \label{fig:lattice}
\end{figure}

The central role played by such local constraints may best be illustrated by spin ice~\cite{Castelnovo12}.
Spin ice is an instance of a classical spin liquid having physical realizations in the rare-earth pyrochlore magnets $\mathrm{Ho_2Ti_2O_7}$~\cite{Harris97} and $\mathrm{Dy_2Ti_2O_7}$~\cite{Bramwell01}.
These materials do not show any long range magnetic order when cooled.
Instead, at low temperatures, the spin configurations satisfy a local constraint known as the ``ice rules''~\cite{Bernal33,Pauling35}, in which two spins point into and two spins point out of each of the tetrahedra which make up the pyrochlore lattice (\emph{cf.}~Fig.~\ref{fig:lattice}).
The number of of spin configurations satisfying this constraint scales exponentially with the size of the system, leading to a ground state ``ice'' manifold with an extensive residual entropy of approximately $\ln(3/2)/2$ per site~\cite{Pauling35,Ramirez99}.

The ice rules constraint also controls spin correlations, which are algebraic \cite{Henley05,Henley10}, and constrains the spin dynamics, thereby giving rise to fractionalized excitations---magnetic monopoles~\cite{Ryzhkin05, Morris09, Jaubert09, Castelnovo12}.
Collectively, these phenomena can be described within the framework of an emergent $\mathrm{U}(1)$ gauge theory, with the ice rules constraint playing the role of a generalized Gauss' law~\cite{Castelnovo08a,Henley10,Castelnovo12}.
As a consequence, the local constraint fully characterizes the resulting classical spin liquid, much as an order parameter might characterize a broken symmetry phase.

It is important to be aware that the presence of a constrained, extensively-degenerate, ground-state manifold, does not preclude symmetry-breaking order.
Examples to the contrary include the order-by-disorder scenario responsible for the hidden quadrupolar~\cite{Chalker92} and octupolar order~\cite{Zhitomirsky08} in the classical Heisenberg antiferromagnet on the kagome lattice, and the ``moment fragmentation'' mechanism~\cite{Brooks-Bartlett14}, seen in the excitations of $\mathrm{Nd_2Zr_2O_7}$~\cite{Petit16,Benton16-PRB94}.
Such situations nevertheless would not elude our machine as it is capable of picking up both order parameters and constraints.

The remainder of this paper is organized as follows.
In Section~\ref{sec:method}, we define the XXZ pyrochlore antiferromagnet and provide an introduction to TK-SVM and spectral graph partitioning.
The principle and the physical relevance of the methods are discussed in detail.

In Section~\ref{sec:graph}, we illustrate the general procedure for using TK-SVM in conjunction with the graph analysis to explore an unknown phase diagram by applying it to the XXZ pyrochlore antiferromagnet.
It is important to note that, in this step, the phase diagram is constructed without relying on physical quantities such as order parameters or the specific heat.

Section~\ref{sec:patterns} is devoted to the interpretation of the machine results.
We present the procedure for extracting the analytical order parameters and emergent local constraints.
The nature of the phases and their characterization will thus be made clear.
In addition, a disorder hierarchy is introduced, which allows us to discuss relations of the phases going beyond the phase diagram.

In Section~\ref{sec:d_functions}, we examine the thermodynamics of the local constraints and the ability of TK-SVM to isolate different phase transitions and crossovers.
There, the behavior of the extracted order parameters also verifies the phase diagram attained from the earlier partition of the graph.

Finally, we conclude in Section~\ref{sec:summary} with an outlook.

\section{Model and methods}
\label{sec:method}

Machine learning techniques are tailored to discover structure in complex data~\cite{BookBishop}.
They provide new tools for the study of many-body problems, in particular when the nature of a system is not evident from known information.
Frustrated magnets are systems of this kind.
Although these systems offer a plethora of novel phases, it is notoriously difficult to identify the right characterization of those phases, especially in the presence of spin liquids or hidden orders.

To apply machine learning methods to frustrated systems, a critical factor is that human can understand the cause of machine's decision and can translate its results to physical language.
Namely, the machine needs to be strongly interpretable.
This is because, for the purpose of understanding a system, we seek an analytical description of its distinctive degrees of freedom, such as order parameters or other characteristics.

Such interpretability would be even more significant when systems feature a complicated phase diagram, which is common in frustrated magnets.
In those cases, aside from learning a numerical label for each phase, we need to interpret the meaning and the relation of those labels.
Furthermore, if a phase involves multiple orders, it is also necessary to identify and extract them individually.

Additional to interpretability, another crucial factor in practical uses is expressiveness of the machine. 
That is the desired machine is expected to be applicable to a large class of systems, with minimal modification to its architecture.
Thus one does not have to delicately design a machine for every single problem.

However, owing to the well-known interpretability-expressibility trade-off,
to integrate the two factors is a very challenging issue.
The tensorial kernel support vector machine (TK-SVM) is a machine that, in the context of studying frustrated systems, reconciles interpretability and expressibility \cite{Greitemann19, Liu19}.
It inherits strong interpretability from standard SVMs, while in the meantime the tensorial kernel is capable of detecting general spin (nematic) orders and, as we shall see in this work, emergent local constraints.

In this section, we introduce the TK-SVM and an associated graph analysis for partitioning phase diagrams.
The use of the latter renders TK-SVM effectively working as an \emph{unsupervised} machine learning method in the sense that we do not rely on prior information of the phase diagram or prior training with data from known phases, although standards SVMs fall into supervised schemes.
We will focus on their physical implications and, for completeness, also briefly review standard SVMs.
A pyrochlore antiferromagnet will be considered as a test model.
The discussion is nevertheless intended to provide general guidance on the use of TK-SVM and is transferable to other spin models.

\subsection{XXZ model}

We consider the XXZ model on the pyrochlore lattice studied in Ref.~\onlinecite{Taillefumier17}.
This model accommodates a multitude of exotic phases, and is, hence, suitable for the purpose of demonstrating the capabilities of TK-SVM.

The Hamiltonian of the XXZ model is given by
\begin{align}
\label{eq:H_XXZ}
H_\textup{XXZ} = \sum_{\langle i,j \rangle} J_{zz} S_{i,z} S_{j,z} -J_{\pm} (S^{+}_i S^{-}_j + S^{-}_i S^{+}_j),
\end{align}
where $\mathbf{S}_i = (S_{i,x}, S_{i,y}, S_{i,z})$ are classical Heisenberg spins of length $\|\mathbf{S}_{i}\|=1$, with $N$ spins in the system, and $S^\pm_i = S_{i,x} \pm iS_{i,y}$.
The spins are expressed in the local frame attached to their sublattice such that the anisotropic $z$-axis corresponds to the local [111] easy axis of the pyrochlore lattice.
The local unit vectors are given in Appendix~\ref{app:latt}.

The spin configurations used as input to the SVM are obtained from classical Monte Carlo simulations of the Hamiltonian $H_\textup{XXZ}$ [Eq.~(\ref{eq:H_XXZ})]. These simulations were carried out for a system of $N=16L^{3}$ spins, where $L^{3}$ is the number of cubic unit cells.
The results presented in this paper used a system of size $L=8$, with $N = 8192$ spins.
(See Appendix~\ref{app:latt} for details of the simulation.)

\subsection{Support vector machines} \label{sec:svm}
Before introducing TK-SVM, we briefly review standard SVMs.
We will focus on the basic notions that are used latter in TK-SVM, and refer to Ref.~\onlinecite{Liu19} for a (minimal) technical review and Refs.~\onlinecite{BookVapnik, Scholkopf00} for comprehensive discussions.

Support vector machines (SVMs) belong to the class of supervised machine learning schemes. They construct a decision boundary, separating two classes of labeled training data~\cite{CortesVapnik95}.
In the context of learning the phase diagram of a frustrated magnet, each class may be a set of spin configurations.

The associated decision function, $d(\mathbf{x})$, gives the oriented distance of an unclassified test sample $\mb{x}$ from the decision boundary,
\begin{align} \label{eq:decfun-kernel}
  d(\mb x) = \sum_k \lambda_k y^{(k)}K(\mb x^{(k)}, \mb x)-\rho,
\end{align}
such that the decision boundary is defined by solutions to $d(\mb{x})=0$.
Here, $\mb{x}^{(k)}$ denotes the $k$-th training sample.
A binary label, $y^{(k)}=\pm 1$, is assigned to each $\mb{x}^{(k)}$.
$K(\mb x^{(k)}, \mb x)$ is a kernel function which (implicitly) maps the samples to an auxiliary space where data are linearly separable.
Hence, the decision boundary is given by a hyperplane in that auxiliary space.
$\lambda_k$ are Lagrange multipliers and quantify the contribution of $\mb{x}^{(k)}$ to the decision function;
samples with $\lambda \neq 0$ are known as \emph{support vectors}. The constant $\rho$ offsets the hyperplane from the origin and is known as the bias parameter. The decision function is fully determined by the set of all $\lambda_k$ and $\rho$. These constitute the output of an SVM.

If solutions do exist, there are typically infinitely many.
To select the optimal solution, a margin of finite width is imposed around the hyperplane such that the margin boundaries on either side are given by $d(\mb{x}) = \pm 1$ and the width of the margin is sought to be maximized, selecting the hyperplane with the most clearance from any samples~\cite{BookVapnik}.
In order to address the common scenario that no solution exists, the margin can be softened by allowing for incursions into the margin and even samples falling onto the wrong side of the decision boundary at a penalty to the optimization objective.
The strength of the regularization is controlled by a parameter $\nu\in [0,1)$ where $\nu = 0$ corresponds to the unregularized ``hard'' margin limit~\cite{Scholkopf00}.
In principle, the choice of the regularization parameter needs to be cross-validated against the training data.

We have studied the influence of the regularization parameter in the context of TK-SVM (\emph{cf.} the supplementary materials to Ref.~\onlinecite{Greitemann19}, as well as Ref.~\onlinecite{Liu19}). We found that a stronger regularization yields higher-quality results for the sake of extracting analytical quantities from the decision function ($\nu=0.5$ is used in Secs.~\ref{sec:patterns} and \ref{sec:d_functions}) but a weaker regularization is the safer choice when probing an unknown phase diagram ($\nu=0.1$ is used in Sec.~\ref{sec:graph}).

\subsection{Tensorial kernel support vector machines} \label{sec:tk-svm}

The applicability of SVM to a given problem crucially depends on the choice of a suitable kernel function.
We introduced a tensorial kernel (TK) which has been shown to be able to identify general tensorial spin orders~\cite{Greitemann19} and allows for the efficient exploration of intricate phase diagrams~\cite{Liu19}.

The training data are given by the full spin configurations, $\mb{x} = \{S_{i,a}\}$ ($i=1,\dots,N; a=x,y,z$).
Each $\mb{x}$ will be assigned a label. However, as has been exploited in Ref.~\onlinecite{Liu19} and will be revisited in Sec.~\ref{sec:partition},
the labeling can be trivial and merely used to distinguish a point from others in the parameter space.
Thus, one does \emph{not} need to rely on information from which phase the spin configurations are sampled; rather, such information will be output by   TK-SVM.

The tensorial kernel is then defined as a quadratic kernel,
\begin{align}
  K(\mb{x}, \mb{x}^\prime) &= \big[\bds{\phi}(\mb{x})\cdot\bds{\phi}(\mb{x}^\prime)\big]^2,\label{eq:kernel}
\end{align}
with respect to transient feature vectors $\bds{\phi}(\mb{x})$.
$\bds{\phi}(\mb{x})$, as given in Eq.~\eqref{eq:phi}, will construct a set of basis tensors that can express general spin orientational orders, including both dipolar and high-rank tensorial ones.
We shall now discuss the physical implications of this kernel and the corresponding TK-SVM decision function.

\subsubsection{Tensorial features}
The feature vector $\bds{\phi}$ is defined as a mapping of $\mb{x}$ to monomials of spin components,
\begin{align} \label{eq:phi}
  \bds{\phi}(\mb{x}) =\{\phi_{\mu}\}= \{ \langle S^{\alpha_1}_{a_1} \dots S^{\alpha_n}_{a_n} \rangle_{\rm cl} \}.
\end{align}
In this mapping the system is partitioned into local clusters of $r$ spins, with spins within a cluster labeled by Greek letters
$\alpha_1, \dots, \alpha_n = 1, 2, \dots,r$.
The monomials of degree $n$ correspond to the elements of the rank-$n$ basis tensors
$\mb{S}^{\alpha_1} \otimes \mb{S}^{\alpha_2} \otimes \cdots \otimes\mb{S}^{\alpha_n}$ from which an order parameter tensor or constraint can be constructed.
$\mu = (\alpha_1, a_1; \dots; \alpha_n, a_n)$
 collects the spin and component indices in a given cluster, $a_1, \dots, a_n = x,y,z$.
Moreover, a lattice average over spin clusters,
$\langle \dots \rangle_{\rm cl}$, is applied to reduce the dimension of the data.
This ansatz assumes that the $r$ spins in the cluster will suffice to express the underlying local orders and constraints,
based on the nature of a local quantity.

The dimension of $\bds{\phi}(\mb{x})$ is determined by the degree ($n$) of the monomials and the size ($r$) of the cluster,
$\mathrm{dim}[\bds{\phi}(\mb{x})] = (3r)^n$.
However, the mapping in Eq.~\eqref{eq:phi} exhibits a large redundancy, since monomials related by a simultaneous swap of the spin and component indices are equivalent, e.g.
$\langle S^{\alpha_1}_{a_1} S^{\alpha_2}_{a_2} \dots \rangle_{\rm cl}
= \langle S^{\alpha_2}_{a_2} S^{\alpha_1}_{a_1} \dots \rangle_{\rm cl}$.
The complexity of the underlying SVM optimization depends only on distinct monomials, whose number is given by the  multinomial coefficient
$\left(\!\!\binom{3r}{n}\!\!\right)  = \binom{3r+n - 1}{n} =
\frac{(3r+n-1)!}{n!(3r-1)!}$,
scaling much slower than
$\mathrm{dim}[\bds{\phi}(\mb{x})]$~\cite[supplementary materials]{Greitemann19}.

Note that both the rank of the basis tensors and the optimal spin cluster do not need to be known from the beginning.
For the former, owing to the presence of the lattice environment, orientational tensors of physical interest are bounded by a rank $n_{\rm max} = 6$ (or even smaller for less symmetric lattices)~\cite{Nissinen16}.
One may successively apply the mapping $\bds{\phi}(\mb{x})$, Eq.~\eqref{eq:phi}, from $n=1$ to $n_\textup{max}$.
Thereby, $d(\mb{x})$ will probe tensorial orders at all relevant ranks.
Furthermore, the choice of the spin cluster can be guided by information on the lattice.
A reasonable trial cluster may consist of a number of lattice unit cells.
It is hence possible to capture composite orders, such as bond and plaquette orders, as well as orders with multiple finite wave vectors.
If the cluster is larger than needed, the TK-SVM will learn a reducible representation of the underlying order parameters and/or constraints where a simplification is usually straightforward.

For the purpose of demonstration, we will consider a tetrahedral cluster of four spins for the pyrochlore XXZ model in Eq.~\eqref{eq:H_XXZ} and apply TK-SVM at ranks $n \leq 4$.
The procedure remains transferable to general classical spin models where larger clusters and higher rank tensors may be needed.

\subsubsection{Coefficient matrix}

Plugging the tensorial kernel Eq.~\eqref{eq:kernel} into the decision function Eq.~\eqref{eq:decfun-kernel}, the TK-SVM decision function is obtained and can be expressed in a quadratic form,
\begin{align} \label{eq:decfun}
   d(\mb{x}) = \sum_{\mu \nu} C_{\mu \nu} \phi_{\mu} (\mb{x}) \phi_{\nu} (\mb{x}) - \rho.
\end{align}
The coefficient matrix,
\begin{align}\label{eq:Cmunu}
  C_{\mu\nu} = \sum_{k} \lambda_k  y^{(k)} \phi_{\mu}(\mb{x}^{(k)})\phi_{\nu} (\mb{x}^{(k)}),
\end{align}
sums over support vectors and will be learned from the SVM optimization.
It identifies the relevant basis tensors and encodes the \emph{analytical} expression  of a potential (hidden) order or emergent constraint.
The indices $\mu$ and $\nu$ each enumerate $n$ spins within a cluster and their components. Formally, $C_{\mu\nu}$ is thus a symmetric $(3r)^n\times(3r)^n$ matrix but the number of independent elements is again much smaller, owing to the aforementioned redundancy.

Without dwelling on technical details, the meaning of $C_{\mu\nu}$ may be illustrated for the simple case where a single magnetic order is present.
$C_{\mu\nu}$ then represents a set of contractions between the relevant basis tensors (rank-$1$ tensors in this example), such that
the quadratic part of the decision function will realize the squared magnitude of the underlying magnetization
up to a linear rescaling.
By extracting $C_{\mu \nu}$ after training, one can infer the analytical expression for the order parameter.

In the general case, where a phase may possess multiple coexisting  orders, including hidden nematic orders, and/or emergent local constraints, $C_{\mu\nu}$ will capture them simultaneously.
Examples for such a case will be seen in Section~\ref{sec:patterns}.
Moreover, regardless of its complexity, $C_{\mu\nu}$ retains its interpretability.
This is a crucial feature of TK-SVM and has been validated against the most complicated rank-$6$ tensorial order~\cite{Greitemann19} and coexisting orders~\cite{Liu19}.

\subsubsection{Bias criterion}
\label{sec:bias_criterion}
In addition to the coefficient matrix, the bias $\rho$ is also determined by the underlying SVM optimization.
Although this parameter merely offsets of the decision function by a constant, it admits a physical interpretation in TK-SVM and serves as an indicator of the presence or absence of a phase transition or crossover.

We again use the example of a simple magnetic order to illustrate this implication, whereas a systematic demonstration for general cases is found in Ref.~\onlinecite{Liu19}.
As discussed above, the quadratic part of the decision function Eq.~\eqref{eq:decfun} produces the squared magnitude of the magnetization which will attain a finite value in the ordered phase but vanish in the disordered phase.
Hence, in the latter case, the decision function is equal to $-\rho$.
As has been pointed out in Sec.~\ref{sec:svm}, the margin boundaries are given by the equations $d(\mathbf{x}) = \pm 1$.
The samples from the disordered phase will fall firmly on the appropriate margin boundary---otherwise the margin would have been unnecessarily narrow (defying the optimization objective) or \emph{all} of the disordered samples would incur a penalty according to the regularization.
As a consequence, spin configurations on the disordered side of the phase boundary correspond to
$d(\mb{x}) = -\rho = \pm 1$.

Therefore, if two sets of data, dubbed $A$ and $B$, are separated by an order-disorder phase transition, apart from the underlying order parameter, SVM will also learn a bias with an ideal value of $\rho = \pm 1$.
Which sign of $\rho$ is realized is a matter of convention that fixes the ``orientation'' of the decision function.
In the following, we use ``$A\,|\,B$'' to denote a classification and work with the convention where $\rho(A\,|\,B) = - \rho(B\,|\,A) = -1$ corresponds to the situation that $A$ ($B$) is in the ordered (disordered) phase.

This ``ideal'' value $\rho = \pm 1$ is derived by assuming that the decision function captures a nontrivial quantity that assumes a nonzero value only in one phase while it averages away in the other, isotropic, phase~\cite{Liu19}.
This remains to be valid for crossovers, as long as
$\langle \sum_{\mu \nu} C_{\mu \nu} \phi_{\mu} \phi_{\nu} \rangle= 0$ for samples in the disordered phase.
Based on this, we can infer the behavior of $\rho$ in other situations, which has been verified empirically.

First, if one phase possesses two or more orders, while a subset of them vanishes when entering the disordered phase and the remaining ones only diminish in magnitude,
$\abs{\rho}$ will typically be slightly larger than unity, owing to a contribution from the difference in magnitude of the persevering orders.
Such behavior can occur when dealing with vestigial orders and partial symmetry breaking.

Further, if the two sets of samples originate from the same phase and, hence, are characterized in the same way, $\rho$ can dramatically exceed unity, $\abs{\rho} \gg 1$.
Nevertheless, in those cases, the sign of $\rho$ retains its physical meaning:
A negative $\rho(A\,|\,B)$ indicates that $A$ is relatively deeper in the ordered phase.

Lastly, $\rho$ can also differ significantly from $\pm 1$ but fall into the interval $(-1,1)$.
This may happen when both sample sets originate from nontrivial phases featuring different characteristics.
In that case, even though $C_{\mu\nu}$ can capture the characteristics of both phases, the sign of $\rho$ will lose its above interpretation.
Namely, the TK-SVM can still identify them as distinct phases, but one cannot interpret their relation in terms of a simple order-disorder transition.

This behavior of the bias can be summarized by the following rules, which have proved themselves useful in probing phase transitions and crossovers,
\begin{align}\label{eq:rho_rules}
	\rho(A\,|\,B) \begin{cases}
		\!\begin{rcases}
		  \gg 1\\
		  \ll -1
		\end{rcases}\quad & \textup{$A$, $B$ in the same phase}, \\
		\approx 1 & \textup{$A$ in the disordered phase}, \\
		\approx -1 & \textup{$B$ in the disordered phase}, \\
		\in (-1,1) & \textup{not directly comparable}.
	\end{cases}
\end{align}
Note that, for the convenience of discussing different disordered phases, such as cooperative paramagnets and trivial paramagnets, the two phases are compared by their level of disorder.
(We shall address this further in Sec.~\ref{sec:hierarchy}.)

\subsection{Partition of the phase diagram} \label{sec:partition}
SVMs are fully capable of classifying multiple sets of data.
A simple yet efficient way to implement SVM multiclassifications is to consider each pair of the sample sets as a binary classification problem and solve them individually~\cite{HsuLin02}.
Further, given the bias criteria summarized in Eq.~\eqref{eq:rho_rules}, we can build a graph from the resulting bias parameters and partition it via a spectral cluster analysis.
This combination of TK-SVM with graph theory yields an efficient scheme capable of constructing phase diagrams without prior knowledge~\cite{Liu19}.

Consider a spin system involving a set of physical parameters like temperature or interactions.
To attain the topology of the phase diagram,
instead of scanning those parameters one by one,
we sample spin configurations from, say, $M$ different points covering the parameter space uniformly and assign them distinct labels.
This constitutes $M$ classes of training samples to which we apply TK-SVM multiclassification.
As a result, for each scrutinized rank $n$ and spin cluster, we will obtain $M(M-1)/2$ decision functions, each yielding a separate bias parameter, corresponding to the binary classifiers between any two classes.
Note that, as part of this procedure, samples are labeled simply according to the parameter point where they originate, \emph{i.e.} the labeling does not involve any information on either the phases themselves or the topology of the phase diagram.

We then consider those parameter points as vertices and introduce edges connecting two vertices based on the bias value of the associated decision function.
This builds up an undirected, simple graph with $M$ vertices and, at most, $M(M-1)/2$ edges.
This graph can subsequently be partitioned by a spectral clustering analysis.
Combining the results of this partitioning at different ranks and spin clusters yields a number of subgraphs which can be interpreted as distinct phases,
and accordingly, the structure of the partitioned graph reflects the topology of the phase diagram of interest.

In Ref.~\cite{Liu19} we have utilized this scheme with unweighted graphs.
That is, an edge has been added between two vertices if the corresponding bias dramatically exceeded unity; otherwise the edge was absent.
Intuitively, this led to graphs where vertices in the same phase were densely intraconnected, while those belonging to different phases were barely connected or remained disconnected.
We showed there that the partition of those graphs could capture symmetry-breaking phases and their transitions remarkably well.

In this work, we extend this scheme to weighted graphs where edges are assigned a weight in the interval $(0,1)$ based on the value of the bias.
Comparing to unweighted graphs, this strategy retains more of the information provided by the biases and, consequently, improves the sensitivity to crossovers which might otherwise be drowned out by phase transitions.
Nonetheless, the choice of the weighting function does not appear to have significant impact in practice as we will discuss in the forthcoming section.

One should note that in the graph analysis outlined above, we do not rely on monitoring single phase transitions or tuning parameters individually.
The topology of the phase diagram is in fact resolved by direct observation of the entire parameter region of interest while simultaneously scrutinizing various potential orders.
Thus, it can be particularly useful when phase diagrams are multidimensional and complex in structure.

\section{Topology of the phase diagram}
\label{sec:graph}

\begin{figure}
  \centering
  \includegraphics{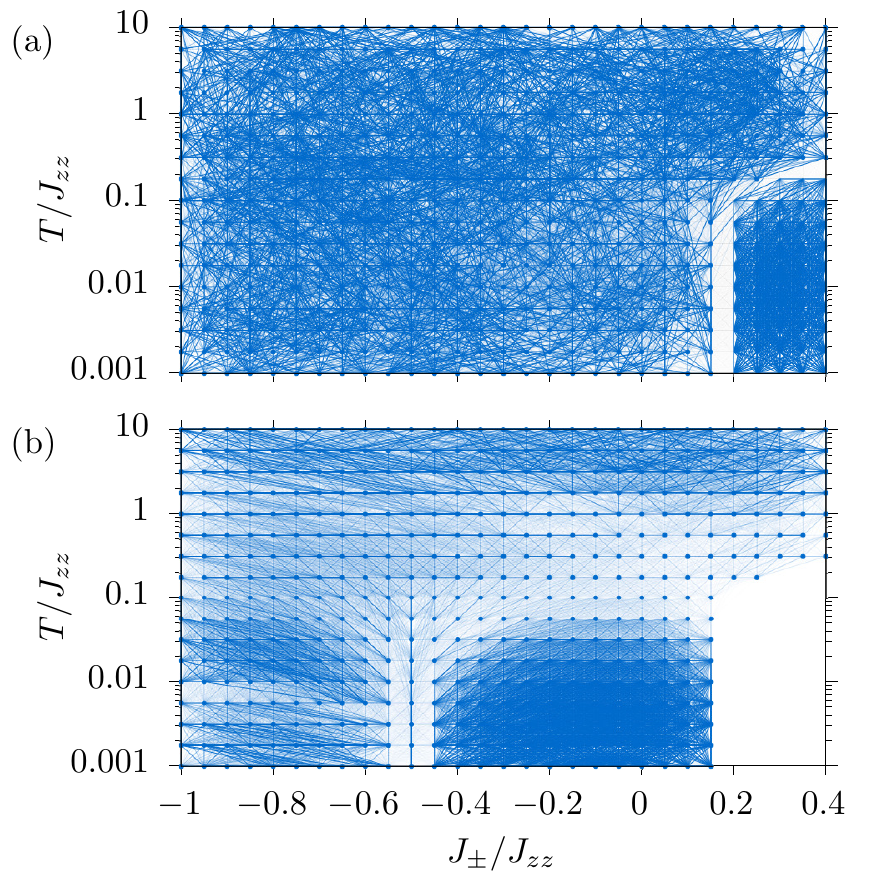}
  \caption{The graphs constructed from the biases between each pair of vertices using the tensorial kernel at rank 1 (a) and rank 2 (b). The opacity of each edge indicates its weight which is in both cases determined using a Lorentzian, see Eq.~\eqref{eq:lorentzian}, with $\rho_c=2$ (a) and $\rho_c=25$ (b), respectively. Here, only edges between (at most) sixth-nearest-neighboring grid points are considered,~\emph{i.e.} long-range edges have been excluded, merely to reduce the visual density of the figure; the subsequent analysis includes long-range edges as well.}
  \label{fig:graph}
\end{figure}

\begin{figure*}
  \centering
  \includegraphics{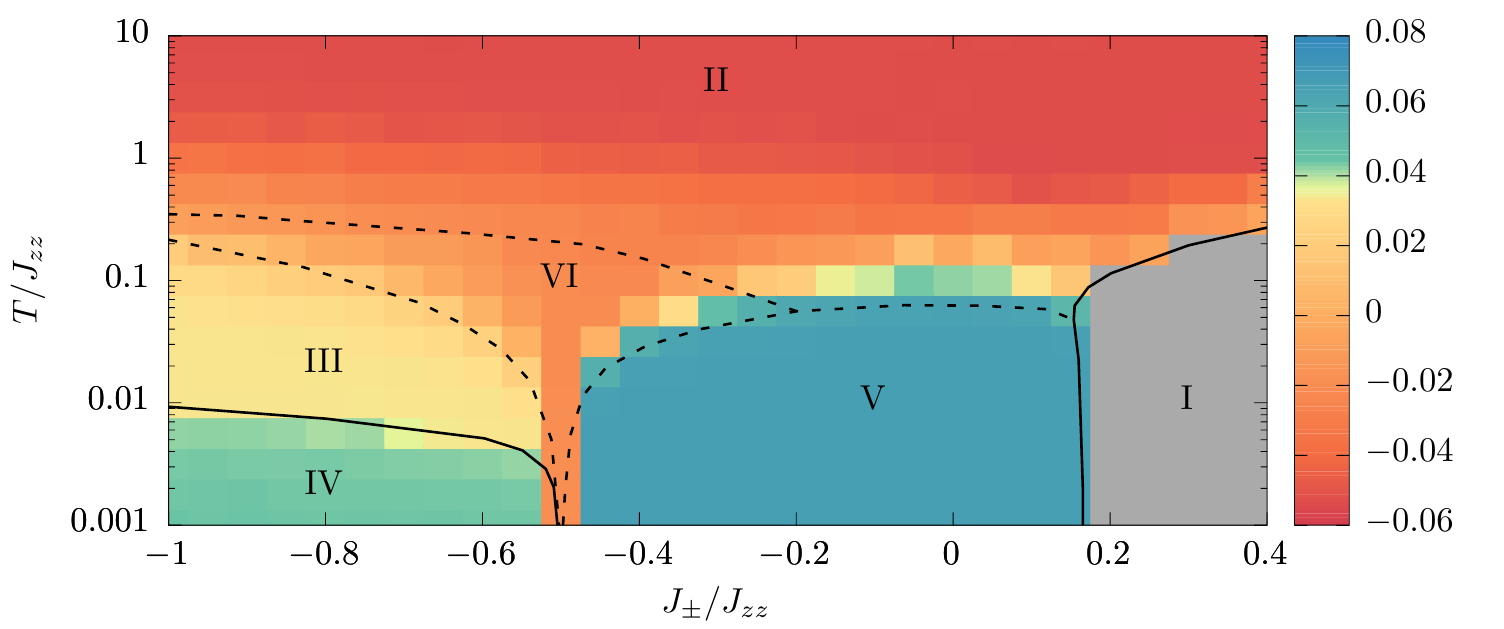}
  \caption{
  Phase diagram of an XXZ pyrochlore anti-ferromagnet, determined by the tensorial kernel support vector machine (TK-SVM).
   The phase diagram is obtained in terms of color-coding the entries of the Fiedler vector.
   Each entry is shown via a pixel and corresponds to a vertex in parameter space.
   The underlying weighted graph, Fig.~\ref{fig:graph} (b), has been constructed from a rank-2 kernel using a Lorentzian weighting function with characteristic bias of $\rho_c=50$ (see Appendix~\ref{app:graph_weights}).
	The grey area ``I'' is classified through a rank-1 TK-SVM analysis, shown in Fig.~\ref{fig:graph} (a), and is omitted from the analysis here.
  For comparison, we include the phase boundaries found in Ref.~\onlinecite{Taillefumier17}. Solid (dashed) lines indicate phase transitions (crossovers).}
  \label{fig:fiedler}
\end{figure*}

\begin{figure}
  \centering
  \includegraphics{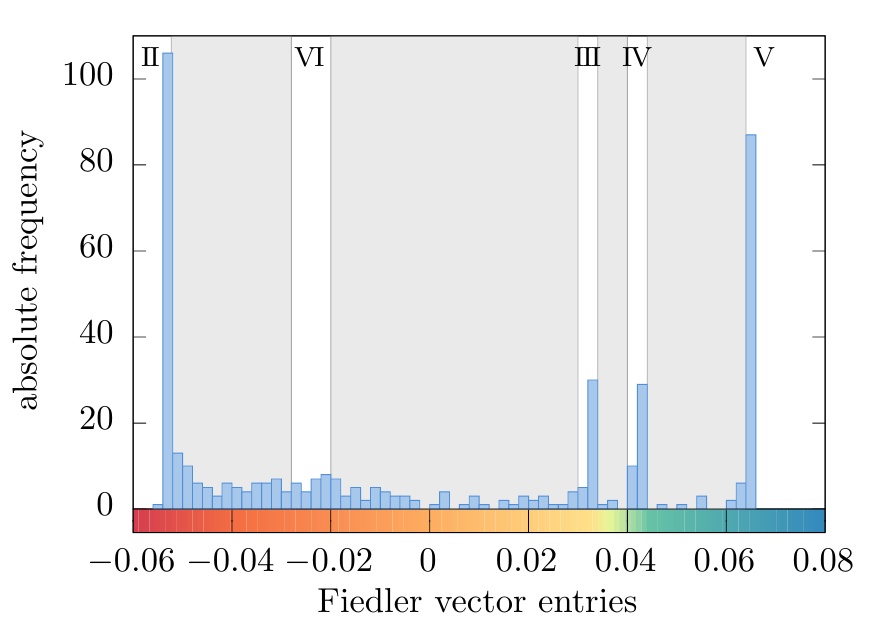}
  \caption{Histogram of the entries of the Fiedler vector shown in Fig.~\ref{fig:fiedler}, alongside the color scale that was used therein. The four regions~II, III, IV, and V can be readily identified as distinct peaks in the histogram. Region~VI is less well-defined, owing to crossovers into other regions. The shaded areas in the background indicate which intervals of Fiedler vector entries were used to guide the pooling of the training data as detailed in Appendix~\ref{app:pooling}.}
  \label{fig:fiedler_histo}
\end{figure}

Having introduced TK-SVM and the partitioning of the phase diagram, we now turn to the concrete application of these techniques to the pyrochlore XXZ model, Eq.~\eqref{eq:H_XXZ}.

To generate the training data, we sample spin configurations on a grid of parameter points $(J_\pm, T)$, covering a region of interest in parameter space uniformly. In the present case, we took 29 equidistant values of $J_\pm$ between $-1$ and $0.4$, as well as 17 logarithmically-spaced temperatures between $0.001$ and $10$.
This choice does not reflect any prior information on the phase diagram, even though a uniform grid is not required per se.

At each of the resulting 493 parameter points, a mere 500 statistically independent spin configurations have been sampled.
These constitute the training data.
Using more samples would further reduce statistical errors~\cite{Greitemann19}; nevertheless, a small number of configurations per parameter point will be seen to already give rise to a remarkably precise phase diagram.
This allows us to explore the parameter space with decent resolution.

The result is a massive multiclassification problem with 493 distinct labels. Solving this using the tensorial kernel at a given rank yields $121,\!278$ decision functions, each trying to distinguish between two parameter points.

We use a Lorentzian weighting function to map the corresponding biases to the edges of the weighted graph.
We justify this choice in Appendix~\ref{app:graph_weights}.

First, we subject the data to an analysis using the rank-1 tensorial kernel, to scrutinize potential dipolar orders. The graph resulting from the corresponding biases is shown in Fig.~\ref{fig:graph}(a).
It is apparent that the graph decomposes into two subgraph components.
Both subgraphs appear to be uniformly intraconnected and do not exhibit any further structure.
This means that, in terms of dipolar orders which are defined on tetrahedral spin clusters, the phase diagram appears split in two.
However, as one cannot foresee or exclude the presence of more complicated orders, it is prudent to analyze both subgraphs further with higher-rank kernels.
In principle, one should also explore the possibility of orders which are defined on larger clusters which we omit here.

In fact, the smaller graph does not exhibit any further structure at ranks 2, 3, or 4 either and is therefore ignored for the remainder of this section.

Hence, we shall analyze the larger subgraph with the rank-2 tensorial kernel next, which detects quadrupolar orders as well as local constraints that can be expressed by terms which are quadratic in the spin components.
The resulting graph is displayed in Fig.~\ref{fig:graph}(b) and exhibits several regions which are strongly intraconnected, corresponding to regimes of congruent nature.
At ranks 3 and 4, no further subdivisions become apparent, indicating that all the relevant phases can be described in terms of quadrupolar order parameters or quadratic constraints.

In order to objectify the identification of these regimes, we perform a spectral clustering analysis of the graph.
This entails the construction of the graph's Laplacian matrix $L=D-A$ and its eigendecomposition. Here, $D$ is the diagonal degree matrix whose diagonal elements hold the sum over the weights of all edges incident on the corresponding vertex, and $A$ is the adjacency matrix whose off-diagonal elements hold the individual weights of the edge connecting the vertices corresponding to its row and column index, respectively.
Given that the graph is connected (which is almost guaranteed when using a continuous weighting function), the Laplacian's smallest non-zero eigenvalue has an associated eigenvector, called Fiedler vector, which can be used to infer an optimal partition of the graph~\cite{Fiedler73, Mohar89}.

The resulting Fiedler vector can be interpreted as the phase diagram, as depicted in Fig.~\ref{fig:fiedler}.
Each of its entries corresponds to a vertex in the parameter space (in other words a pixel in Fig.~\ref{fig:fiedler}).
We observe that those entries attain distinct values within strongly intraconnected regions which are relatively constant throughout.
In order to demonstrate the usage of TK-SVM in the absence of prior information, we label these regions anonymously by the roman numerals given in the figure for now.
Their nature will become clear once we characterize them in the next section.

However, already at this point can we note the remarkable agreement of this phase diagram with the phase transitions and crossovers found in Fig.~1 of Ref.~\onlinecite{Taillefumier17} which are superimposed on Fig.~\ref{fig:fiedler} for reference.
Also note that the boundary between regions~III and IV is very sharp, whereas the distinction between some other regions is more gradual.
It will be confirmed in Section~\ref{sec:d_functions}, where we examine the corresponding order parameters,
that the former represents a phase transition, while the latter marks the region influenced by a crossover.
Therefore, in addition to learning the topology of the phase diagram, the graph analysis can also distinguish crossovers from phase transitions.

The above partitioning is also reflected by the histogram of the Fiedler vector entries which is presented in Fig.~\ref{fig:fiedler_histo}.
The regions labeled II, III, IV, and V manifest themselves as distinct peaks in the histogram. In particular, regions~III and IV lie closely together, but can be distinguished quite clearly which justifies our choice of a color scale with a large gradient in the vicinity.
On the other hand, the region which we labeled VI is less well-defined.
We attribute this to the fact that it corresponds to an intermediate regime, both in coupling $J_\pm$ and temperature $T$, connected via crossovers to the surrounding regions with drastically different entries in the Fiedler vector.


\section{Characterizing the phases} \label{sec:patterns}
Having established the topology of the phase diagram, the next step is to understand the nature of the phases and their phase transitions and crossovers.
This will be done by employing a reduced multiclassification  based on the learned phase diagram.
From the coefficient matrices of this multiclassification, we extract the analytical order parameters and constraints,
and from the bias parameters we infer a hierarchy between the phases.

To be concrete, we will merge samples belonging to the same region in Fig.~\ref{fig:fiedler} and relabel them accordingly.
To avoid potential misclassification and statistical errors, samples in the (suspected) crossover regions, which cannot be attributed to a single phase unambiguously, will be discarded.
The systematic procedure of this merger is detailed in Appendix~\ref{app:pooling}.

In addition to the merged samples, we supplement the training data with a number of fictitious configurations, consisting of random spins which are independently and isotropically sampled from the unit sphere.
These fictitious configurations mimic states at infinite temperature and serve as a control group which turns out to be useful for interpreting both the coefficient matrices and the behavior of the bias parameters.
We label them as ``$T_\infty$'' in the following discussion.

This sets up a reduced multiclassification problem.
For each kernel of rank $n$, the solution leads to
$q_n(q_n+1)/2$ binary classifiers,
where $q_n$ denotes the number of phases identified at any given rank (here, $q_{1} = 2$, $q_{2} = 5$).
The first $q_n$ of the classifiers involve the control group and encode the characteristics distinguishing each phase from featureless $T_{\infty}$-states.
As we shall see in this section, the interpretation of these $q_n$ classifiers will give sufficient information for understanding the phase diagram.
The remaining $q_n(q_n-1)/2$ classifiers emphasize the distinction between any two phases in the real data, hence providing a straightforward way to identify quantities that are responsible for phase transitions or crossovers.

In the remainder of this section, we will first discuss how the bias parameter can be used to establish a hierarchical relation between phases, and then detail the procedure of extracting analytical quantities from the coefficient matrices.

\begin{table}
  \centering
  \subfloat[Phase~I is compared with the remainder of the parameter space and the control group using the rank-1 kernel.]{
  \begin{tabular*}{0.475\textwidth}{@{\extracolsep{\fill}}c*{2}{D{.}{.}{2.4}}}
  \toprule\addlinespace[-0.05em]
  \midrule
    & \multicolumn{2}{c}{$\rho$}\\\cmidrule(lr){2-3}
	& \multicolumn{1}{c}{I} & \multicolumn{1}{c}{II--VI}\\\midrule
  $T_\infty$ & 1.0006 & 1.4874\\
  I & & -1.0004\\
  \midrule\addlinespace[-0.05em]
  \bottomrule
  \end{tabular*}}

  \subfloat[Phases~II--VI are compared amongst each other and with the control group using the rank-2 kernel.]{
  \begin{tabular*}{0.475\textwidth}{@{\extracolsep{\fill}}c*{5}{D{.}{.}{2.4}}}
  \toprule\addlinespace[-0.05em]
  \midrule
    & \multicolumn{5}{c}{$\rho$}\\\cmidrule(lr){2-6}
	& \multicolumn{1}{c}{II} & \multicolumn{1}{c}{III} & \multicolumn{1}{c}{IV} & \multicolumn{1}{c}{V} & \multicolumn{1}{c}{VI}\\\midrule
  $T_\infty$ & 4.586 & 1.012 & 1.009 & 1.004 & 1.025\\
  II & & 1.026 & 1.016 & 1.012 & 1.097\\
  III & & & 1.336 & 0.534 & -1.220\\
  IV & & & & 0.383 & -1.134\\
  V & & & & & -1.028\\
  \midrule\addlinespace[-0.05em]
  \bottomrule
  \end{tabular*}}
  \caption{Biases $\rho$ of decision functions at different ranks between the data which are labeled according to the phase diagram partitioning obtained in Sec.~\ref{sec:graph}.
  In both cases, we additionally include a fictitious data set of independently distributed isotropic spin configurations, referred to as ``$T_\infty$'', as a control group.
  These biases are to be interpreted according to the rules laid out in Sec.~\ref{sec:bias_criterion} and summarized by Eq.~\eqref{eq:rho_rules}.}
  \label{tab:rho}
\end{table}

\subsection{The hierarchy of disorder}\label{sec:hierarchy}
In Section~\ref{sec:method} we pointed out that the bias parameter $\rho$ is oriented. During the construction of the phase diagram Fig.~\ref{fig:fiedler}, only the magnitude of $\rho$ was used.
We now take the orientation of the biases in the reduced multiclassification into account, and analyze the relation between the phases.

Note that the set of biases involve classifications between all pairings of any two phases.
We can infer a global hierarchy between the phases from them, going beyond the topology of the phase diagram.
This introduces the notion of a ``hierarchy of disorder'' which addresses order-to-disorder transitions and crossovers on an equal footing.
A trivial paramagnet has the most disorder; symmetry-breaking phases, where spins align along common directions, are in the opposite limit;
phases of constrained dynamics that do not break any symmetries reside in the middle.

The bias values resulting from the reduced multiclassification problems are tabulated in Table~\ref{tab:rho}.
According to the rules set out in Eq.~\eqref{eq:rho_rules}, we can infer that phase~I has the least disorder as the corresponding bias $\rho(\textup{I}\,|\,\textup{II--VI})$ is approximately equal to $-1$.
In contrast, phase~II is entirely disordered and can be identified as the paramagnetic phase.
Phase~VI is the second most disordered phase, as only
$\rho((T_\infty,\textup{II})\,|\,\textup{VI}) \approx +1$.
Also note that both phases~III and IV do not directly compare to phase~V, as indicated by their biases in $(-1, 1)$.

With a little further analysis, we can summarize the Table~\ref{tab:rho} by the following hierarchy,
\begin{equation}\label{eq:hierarchy}
\begin{tikzcd}[column sep=tiny, row sep=small]
& & & {\rm III} \ar[rr] & & {\rm IV} \ar[dr] & \\
\big(T_\infty, {\rm II}\big) \ar[rr] & & {\rm VI} \ar[ur] \ar[drr] & & & & {\rm I}. \\
& & & & {\rm V} \ar[urr]
\end{tikzcd}
\end{equation}
These relations are confirmed by the analytical characterizations of the phases which we will discuss in forthcoming subsections.
We will see that phase~VI is a classical $\mathrm O(3)$ spin liquid characterized by an isotropic local constraint.
Phases~III and V feature constraints in the easy-plane and easy-axis, respectively, and therefore experience less disorder.
In addition, phases~I and IV, which are not adjacent in the phase diagram, are two spontaneously symmetry-breaking phases, while the former breaks more symmetry and, hence, comes last in the hierarchy.

\subsection{Identification of broken symmetries} \label{sec:ops}

We now extract the analytical characterization of the phases from the coefficient matrices, $C_{\mu\nu}$ in Eq.~\eqref{eq:decfun}.
We focus here on the local orders in phases~I through IV, but save the discussion of their emergent constraints for the next subsection.


\subsubsection{Rank-1 order}
Let us first focus on results learned with the rank-$1$ decision function.
At this rank, the graph in Fig.~\ref{fig:graph}(a) contains only two disconnected components.
Moreover, the $C_{\mu\nu}$ matrix learned to distinguish the two subgraphs appears identical to $C_{\mu\nu}(\textup{I}\,|\,T_\infty)$,
suggesting one dipolar order is detected in phase~I.

\begin{figure}
  \centering
  \includegraphics{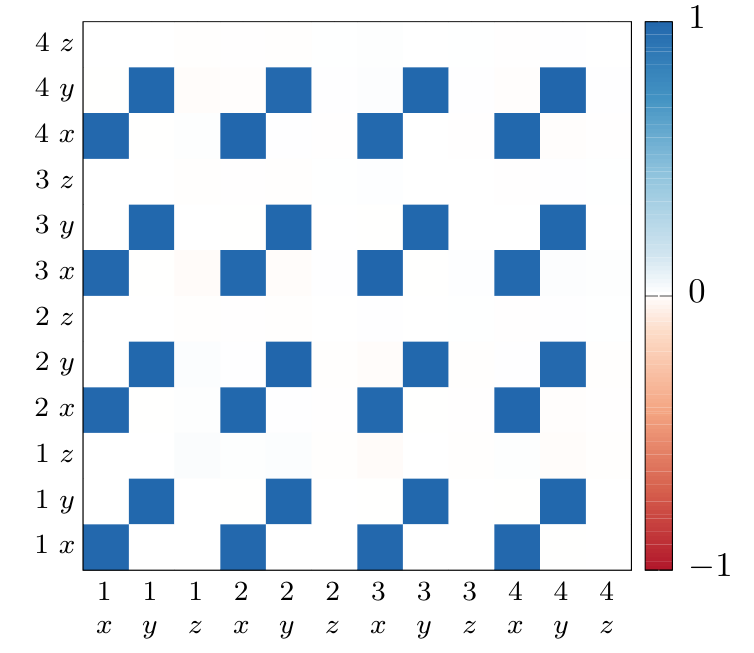}
  \caption{The coefficient matrix $C_{\mu\nu}(\textup{I}\,|\,T_\infty)$, characterizing phase~I against the control group, obtained from TK-SVM at rank 1. The axes are labeled according to sublattice index, $\alpha=1,2,3,4$, and spatial component, $a=x,y,z$, in lexicographical order.
   The repeating $3\times 3$ motif, see Eq.~\eqref{eq:block_AF}, reveals phase~I as an easy-plane antiferromagnet.
   As the spins are defined in the local frames attached to each sublattice in the Hamiltonian Eq.~\eqref{eq:H_XXZ}, the four spins in a tetrahedron do not align with each other.
   }
  \label{fig:AF_pattern}
\end{figure}

The corresponding $12\times 12$ matrix $C_{\mu\nu}$ is shown in Fig.~\ref{fig:AF_pattern}.
Following the definition in Eq.~\eqref{eq:Cmunu}, it is expanded by rank-$1$ basis functions $S^\alpha_a$,
where $\alpha = 1,2,3,4$ and $a = x,y,z$.
The nonvanishing entries identify which of these contribute to the underlying order parameter.

Furthermore, Fig.~\ref{fig:AF_pattern} shows a periodic structure of $3 \times 3$ blocks.
This indicates that $C_{\mu\nu}(\textup{I}\,|\,T_\infty)$ learned with the four-spin tetrahedral cluster is reducible, because the order parameter can be inferred from a single block.
According to their spin indices, we can assign a coordinate $[\alpha, \alpha']$ to each block, and express them by submatrices
$\mathcal{B}^{\alpha \alpha'}(\textup{I}\,|\,T_\infty)$,
\begin{align} \label{eq:block_AF}
	\mathcal{B}^{\alpha \alpha'}_{aa'}(\textup{I}\,|\,T_\infty) =
\raisebox{-22pt}{\hbox{
\begin{tikzpicture}[gnuplot]
\gpcolor{color=gp lt color border}
\gpsetlinetype{gp lt border}
\gpsetdashtype{gp dt solid}
\gpsetlinewidth{1.00}
\draw[gp path] (0.000,0.267)--(0.180,0.267);
\draw[gp path] (0.999,0.267)--(0.819,0.267);
\node[gp node right] at (0.073,0.267) {$x$};
\draw[gp path] (0.000,0.600)--(0.180,0.600);
\draw[gp path] (0.999,0.600)--(0.819,0.600);
\node[gp node right] at (0.073,0.600) {$y$};
\draw[gp path] (0.000,0.933)--(0.180,0.933);
\draw[gp path] (0.999,0.933)--(0.819,0.933);
\node[gp node right] at (0.073,0.933) {$z$};
\draw[gp path] (0.167,0.100)--(0.167,0.280);
\draw[gp path] (0.167,1.099)--(0.167,0.919);
\node[gp node center] at (0.167,-0.024) {$x$};
\draw[gp path] (0.500,0.100)--(0.500,0.280);
\draw[gp path] (0.500,1.099)--(0.500,0.919);
\node[gp node center] at (0.500,-0.024) {$y$};
\draw[gp path] (0.833,0.100)--(0.833,0.280);
\draw[gp path] (0.833,1.099)--(0.833,0.919);
\node[gp node center] at (0.833,-0.024) {$z$};
\draw[gp path] (0.000,1.099)--(0.000,0.100)--(0.999,0.100)--(0.999,1.099)--cycle;
\gpfill{rgb color={1.000,1.000,1.000}} (0.000,0.766)--(0.333,0.766)--(0.333,1.099)--(0.000,1.099)--cycle;
\gpfill{rgb color={1.000,1.000,1.000}} (0.333,0.766)--(0.666,0.766)--(0.666,1.099)--(0.333,1.099)--cycle;
\gpfill{rgb color={1.000,1.000,1.000}} (0.666,0.766)--(0.999,0.766)--(0.999,1.099)--(0.666,1.099)--cycle;
\gpfill{rgb color={1.000,1.000,1.000}} (0.000,0.433)--(0.333,0.433)--(0.333,0.766)--(0.000,0.766)--cycle;
\gpfill{rgb color={0.000,0.000,0.000}} (0.333,0.433)--(0.666,0.433)--(0.666,0.766)--(0.333,0.766)--cycle;
\gpfill{rgb color={1.000,1.000,1.000}} (0.666,0.433)--(0.999,0.433)--(0.999,0.766)--(0.666,0.766)--cycle;
\gpfill{rgb color={0.000,0.000,0.000}} (0.000,0.100)--(0.333,0.100)--(0.333,0.433)--(0.000,0.433)--cycle;
\gpfill{rgb color={1.000,1.000,1.000}} (0.333,0.100)--(0.666,0.100)--(0.666,0.433)--(0.333,0.433)--cycle;
\gpfill{rgb color={1.000,1.000,1.000}} (0.666,0.100)--(0.999,0.100)--(0.999,0.433)--(0.666,0.433)--cycle;
\draw[gp path] (0.000,1.099)--(0.000,0.100)--(0.999,0.100)--(0.999,1.099)--cycle;
\gpdefrectangularnode{gp plot 1}{\pgfpoint{0.000cm}{0.100cm}}{\pgfpoint{0.999cm}{1.099cm}}
\end{tikzpicture}

}}
= \delta_{aa'}(1-\delta_{a,z}),
\end{align}
where only $a,a' = x,y$ components are relevant.

Substituting
$C_{\mu\nu}(\textup{I}\,|\,T_\infty)
 = \{\mathcal{B}^{\alpha \alpha'}\}$
back into the decision function, Eq.~\eqref{eq:decfun}, we obtain
\begin{align}
	d(\mb{x}) \sim \frac{1}{N^2} \sum_i \left\langle(S_{i,x})^2 + (S_{i,y})^2\right\rangle_\textup{cl}
	= \left\langle\|\mb{M}_{\perp}\|^2\right\rangle_\textup{cl}.
\end{align}
One realizes that the interpretation of coefficient matrix here leads to nothing but easy-plane magnetization
$\mb{M}_{\perp}= \frac{1}{N} \sum_i (S_x, S_y, 0)^T$.
In addition, since the spins are defined by sublattice coordinates, we can conclude phase~I to be an easy-plane antiferromagnetic (AFM) phase.

\subsubsection{Rank-2 orders}
The coefficient matrices learned with higher-rank kernels are interpreted in the same spirit as in the rank-$1$ case.
Namely, we identify important basis functions in $\phi(\mb{x})$ and their weights entering $C_{\mu\nu}$.
Moreover, $C_{\mu\nu}$ can be divided into small blocks,
and it is often sufficient to examine a subset of those blocks and their global structure.
At rank $2$, $C_{\mu\nu}$ is expanded by  $S^{\alpha}_a S^{\beta}_b$ and $S^{\alpha'}_{a'} S^{\beta'}_{b'}$,
and the small blocks are again identified by the spin indices $[\alpha \beta, \alpha' \beta']$.

We observe that the coefficient matrix $C_{\mu\nu}(\textup{II}\,|\,T_\infty)$, which is supposed to learn a quantity to distinguish phase~II from the fictitious $T_\infty$ data, exhibits only noise.
This reflects the result that phase~II is a trivial paramagnet (PM) which is consistent with the result obtained through the bias criterion.
In contrast, the coefficient matrices
$C_{\mu\nu}(\textup{III}\,|\,T_\infty)$ and $C_{\mu\nu}(\textup{IV}\,|\,T_\infty)$
show a regular and robust pattern, indicating that nontrivial features have been detected.

\begin{figure}
	\centering
	\includegraphics{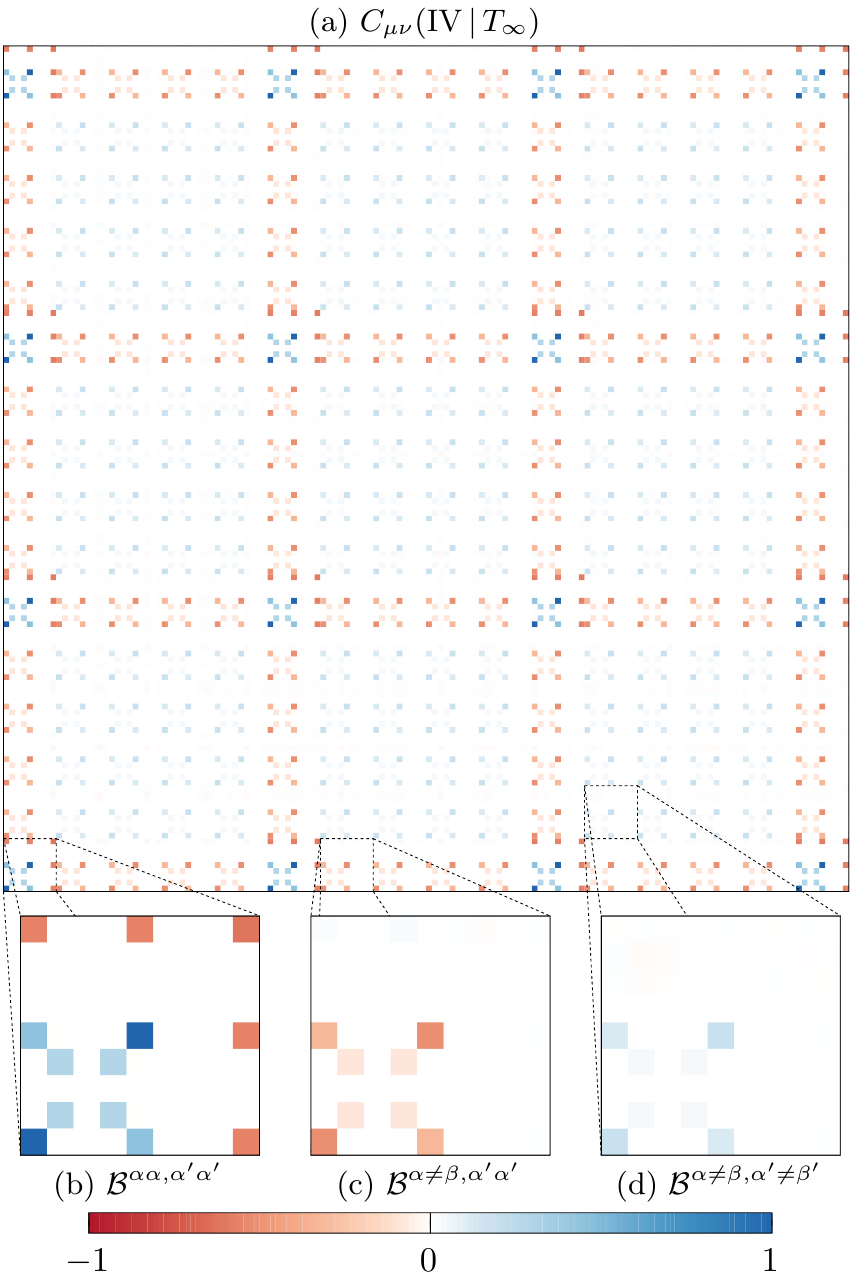}
	\caption{(a) The $144\times 144$ coefficient matrix $C_{\mu\nu}(\textup{IV}\,|\,T_\infty)$ obtained from TK-SVM at rank 2.
	The axes iterate over sublattice indices $\alpha, \beta$ and spatial components $a,b$ such that tuples $(\alpha,\beta,a,b)$ are lexicographically sorted.
	Panels (b) through (d) zoom in to various types of $9\times 9$ blocks of fixed sublattice indices: (b) ``on-site'', (c) ``cross'', and (d) ``bond''.
	The blocks exhibit similar patterns; (c) and (d) differ only by an overall factor and relate to the $C_{2h}$ order parameter; the on-site block (b) additionally contains an equivalent contribution due to the spin normalization.
	Phase~IV is thus an unconventional biaxial spin nematic.}
	\label{fig:SN}
\end{figure}

The pattern of $C_{\mu\nu}(\textup{IV}\,|\,T_\infty)$ is given in Fig.~\ref{fig:SN}(a).
Indeed, one notices a structure of blocks of $9 \times 9$ elements.
Furthermore, these blocks can be classified into three types:
16 ``on-site'' blocks ($[\alpha \alpha, \alpha' \alpha']$),
96 site-bond ``cross'' blocks ($[\alpha \alpha, \alpha' \neq \beta']$
or $[\alpha \neq \beta, \alpha'  \alpha']$),
and 144 ``bond'' blocks ($[\alpha \neq \beta, \alpha' \neq \beta']$), respectively.
Instances of each type are magnified in panels (b) through (d) of Fig.~\ref{fig:SN}.

The structure of these blocks is encompassed by the following submatrix,
\begin{align} \label{eq:block_C2h}
	\mathcal{B}^{\alpha\beta,\alpha' \beta'} (\textup{IV}\,|\,T_\infty) =
	\begin{bmatrix}
		A & & & & A  & \qquad\quad\qquad & B \\
		& & & & & &  \\
		& & & & & &  \\
		& & & & & &  \\
		D & & & & C &  & A \\
		& E & & E & & &  \\
		& & & & & &  \\
		& E & & E & & &  \\
		C & & & & D &  & A
	\end{bmatrix},
\end{align}
where vanishing entries are omitted, and the values of variables $A$ through $E$ can be read off from the patterns.
Moreover, it turns out that it is sufficient to infer local orders in the phase~IV (and also the phase~III) by examining a single block, while the relative strength of different blocks encode an emergent constraint which will be discussed later.

The pattern can be related to three quadrupolar ordering components (two of which occur dependent upon each other; see below):
\begin{align}
	Q^{\alpha\beta}_{x^2+y^2} &\coloneqq S^\alpha_x S^\beta_x + S^\alpha_y S^\beta_y, \label{eq:op_Dinfh_xy}\\
	Q^{\alpha\beta}_{x^2 - y^2} &\coloneqq  S^\alpha_x S^\beta_x - S^\alpha_y S^\beta_y, \label{eq:op_D2h_xy}\\
	Q^{\alpha\beta}_{xy+yx} &\coloneqq S^\alpha_x S^\beta_y + S^\alpha_y S^\beta_x, \label{eq:op_C2h_xy}
\end{align}
whereas the remaining components, $Q^{\alpha\beta}_{z^2}$, $Q^{\alpha\beta}_{yz+zy}$, and $Q^{\alpha\beta}_{zx+xz}$, were not found to be relevant here.
Additionally, within the on-site blocks (Fig.~\ref{fig:SN}(b)), the intrinsic normalization constraint
\begin{align}
	Q^{\alpha\alpha}_{x^2+y^2+z^2} &\coloneqq (S^\alpha_x)^2 + (S^\alpha_y)^2 + (S^\alpha_z)^2 \equiv 1\label{eq:normalization}
\end{align}
contributes a constant. (Its appearance is physically irrelevant; examples of such a ``self-contraction'' have previously been discussed in Ref.~\onlinecite{Greitemann19}.)
Each of the ordering components contributes with a weight $p[Q_\bullet]$ to the makeup of Eq.~\eqref{eq:block_C2h}. The systematic inference of these weights from the coefficient matrix is detailed in Appendix~\ref{app:rank2_op}.

The physical meaning of these ordering components is transparent.
$Q_{x^2 + y^2}$ reflects the anisotropic interaction in the XXZ Hamiltonian Eq.~\eqref{eq:H_XXZ}.
$Q_{xy+yx}$ and $Q_{x^2 - y^2}$ form the hidden order recently discovered in Ref.~\onlinecite{Taillefumier17}.
We observe that they occur with equal weight, $p[Q_{xy+yx}]=p[Q_{x^2 - y^2}]$ (see Table~\ref{tab:components}).
This is consistent with the fact that---given the presence of $Q_{x^2 + y^2}$---$Q_{xy+yx}$ and $Q_{x^2 - y^2}$ define the $C_{2h}$ group polynomial~\cite{Nissinen16, Michel01}.

Therefore, in addition to confirming the findings of Ref.~\onlinecite{Taillefumier17}, the present results also suggest that phase~IV possesses a $C_{2h}$ order.
Its characterization consists of two fluctuating fields, a biaxial one
\begin{align} \label{eq:C2h_op}
	\mathbf{Q}^\mathrm{B}_{C_{2h}} =
	\bigg\langle \left(
	\begin{matrix}
		Q_{xy+yx} \\
		Q_{x^2 - y^2}
	\end{matrix}
	 \right) \bigg\rangle_\textup{cl},
\end{align}
where the components are subject to the relation
\begin{align}
  Q^{\alpha\alpha}_{x^2+y^2}Q^{\beta\beta}_{x^2+y^2} &= (Q^{\alpha\beta}_{xy+yx})^2 + (Q^{\alpha\beta}_{x^2-y^2})^2,\label{eq:component_dependence}
\end{align}
and a uniaxial one
$Q^\mathrm{U}_{C_{2h}} = \langle Q_{x^2 + y^2} \rangle_\textup{cl}$.
The former defines a spontaneously symmetry-breaking order in the easy-plane, while the latter distinguishes it from a $C_{2}$ phase which hosts the same biaxial order~\cite{Nissinen16, Michel01}.

Following the terminology of liquid crystal physics ~\cite{BookdeGennes}, phase~IV may be called a biaxial spin nematic (BSN) phase.
However, in contrast to the well known $D_{2h}$ biaxial phase,
which breaks only rotational symmetries,
this $C_{2h}$ phase also spontaneously breaks mirror symmetries $\sigma_{xz}$ and $\sigma_{yz}$ of the XXZ Hamiltonian Eq.~\eqref{eq:H_XXZ},
hence it is an unconventional biaxial nematic.

\begin{figure}
	\centering
	\includegraphics{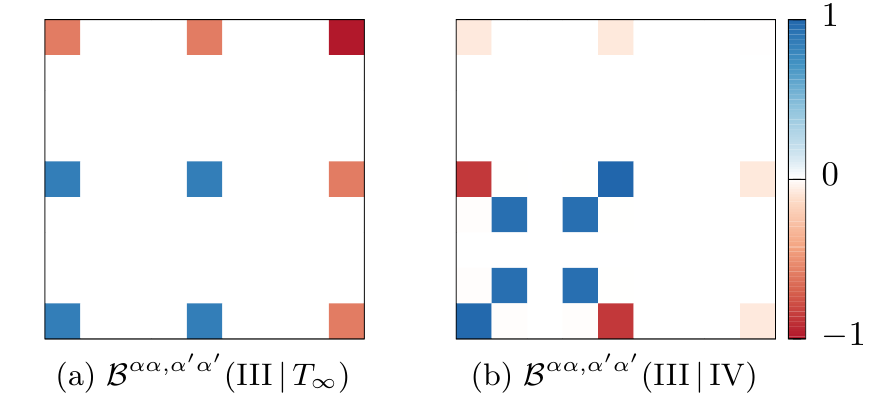}
	\caption{The on-site $9\times 9$ block of the coefficient matrix discerning phase~III from the $T_\infty$ control group~(a) and from phase~IV~(b), respectively, as obtained from the TK-SVM at rank 2.
	The axes iterate over component indices $a,b$ in lexicographical order.
	Both on-site blocks exhibit a redundant contribution due to the spin normalization. Apart from that, (a)~corresponds to the $Q_{x^2+y^2}$ ordering component, revealing phase~III as a uniaxial nematic (without spontaneous symmetry breaking); pattern~(b) corresponds to the $Q_{xy+yx}$ and $Q_{x^2-y^2}$ components, consistent with the fact that phase~IV additionally spontaneously breaks the uniaxial symmetry down to a biaxial one.}
	\label{fig:SL}
\end{figure}

Phase~III is another phase that possesses nonvanishing quadrupolar moments.
The coefficient matrix $C_{\mu\nu}(\textup{III}\,|\,T_\infty)$ displays a similar global structure as $C_{\mu\nu}(\textup{IV}\,|\,T_\infty)$, but a more simple pattern within the small blocks.
A representative block,
$\mathcal{B}^{\alpha\alpha,\alpha' \alpha'} (\textup{III}\,|\,T_{\infty})$,
is provided in Fig.~\ref{fig:SL}(a).
It corresponds to a situation where $C= D$ and $E = 0$ in Eq.~\eqref{eq:block_C2h}.
As a consequence, only the quadrupolar component $Q_{x^2 + y^2}$ is relevant.
However, we note that $Q_{x^2 + y^2}$ is not a spontaneously symmetry-breaking order since the XXZ Hamiltonian explicitly breaks the spin $\mathrm{O}(3)$ symmetry down to the infinite dihedral group $D_{\infty h}$.

Moreover, it is clear that the biaxial order parameter $\mathbf{Q}^\mathrm{B}_{C_{2h}}$ is responsible for the transition between phases~III and IV.
One expects that they will be stressed in the coefficient matrix $C_{\mu\nu}(\textup{III}\,|\,\textup{IV})$.
Indeed, as shown in Fig.~\ref{fig:SL}(b), these components dominate the block $\mathcal{B}^{\alpha\alpha,\alpha' \alpha'} (\textup{III}\,|\,\textup{IV})$, in comparison to
$\mathcal{B}^{\alpha\alpha,\alpha' \alpha'} (\textup{IV}\,|\,T_\infty)$
in Fig.~\ref{fig:SN}(b).

\subsection{Identification of emergent constraints} \label{sec:constraints}
The example of spin ice in the introduction demonstrates that an emergent constraint can dramatically influence correlations and dynamics in the system.
In fact many interesting phases in frustrated systems feature emergent constraints.

Their identification can be far from trivial in the absence of generic tools as they are not obvious from the Hamiltonian.
In special cases, such as the pyrochlore lattice~\cite{Benton16,Yan17,Taillefumier17,Yan19}, water ice~\cite{Benton16-PRB93}, or the kagome lattice~\cite{Essafi16},
obtaining them relies on systematic calculations using group theory (decomposition in terms of irreducible representations).
These calculations are specific to corner-sharing geometries.

In this subsection, we will demonstrate the procedure for deriving emergent constraints from the coefficient matrices.
We will encounter situations in which a phase is defined exclusively by an emergent constraint, and those in which a constraint coexists with symmetry-breaking orders.

\subsubsection{Ice rule}
We first examine the simplest such instance, phase~V.
The procedure is similar to that for analyzing local order in the preceding subsection, but here we shall also evaluate relative weights of different blocks.

\begin{figure}
  \centering
  \includegraphics{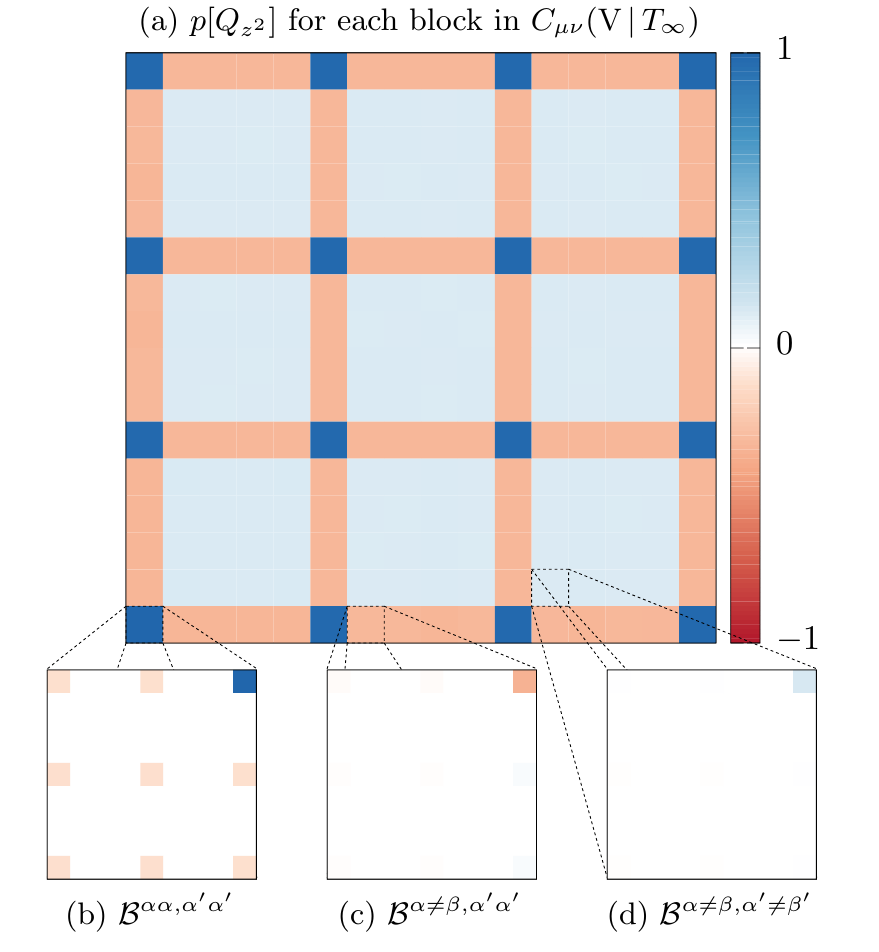}
  \caption{(a) The $16\times 16$ reduced coefficient matrix where each $9\times 9$ block in $C_{\mu\nu}(\textup{V}\,|\,T_\infty)$ is replaced by the weight of the $Q_{z^2}$ ordering component, as obtained through the procedure outlined in Appendix~\ref{app:rank2_op} from the TK-SVM coefficient matrix at rank 2.
	The axes iterate over sublattice indices $\alpha, \beta$ in lexicographical order.
	(b)-(d) Zoom in to various types of $9\times 9$ blocks of fixed sublattice indices: (b) ``on-site,'' (c) ``cross,'' and (d) ``bond.''
	From the relative strength of the $Q_{z^2}$ components in each type of block, one can infer the ice rule.}
  \label{fig:SI-PM}
\end{figure}

To emphasize those weights, in Fig.~\ref{fig:SI-PM}(a) we show a reduced form of the coefficient matrix $C_{\mu\nu}(\textup{V}\,|\,T_{\infty})$.
Each pixel of this reduced matrix now corresponds to a $9\times 9$ block in the full $C_{\mu\nu}$ matrix, while the value of the pixel is given by the weight of the $Q_{z^2}$ ordering component.
One notices that there are three different weights, corresponding to the
site-site ($\mathcal{B}^{\alpha \alpha,\alpha' \alpha'}$),
site-bond cross
($\mathcal{B}^{\alpha \alpha,\alpha' \neq \beta'}$,
$\mathcal{B}^{\alpha \neq \beta,\alpha' \alpha'}$)
and bond-bond
($\mathcal{B}^{\alpha \neq \beta,\alpha' \neq \beta'}$)
blocks in the full $C_{\mu\nu}$, respectively.
Details of these blocks are also provided in Figs. (b)--(d).

The site-site block in Fig.~\ref{fig:SI-PM}(b) can be expressed as
\begin{align} \label{eq:SI-block_1111}
	\mathcal{B}^{\alpha \alpha,\alpha' \alpha'} &=
	p[Q_{x^2+y^2+z^2}]
	\begin{bmatrix}
 	1 & & 1 & & 1\\
 	& & & & \\
 	1 & & 1 & & 1\\
 	& & & & \\
 	1 & & 1 & & 1
	\end{bmatrix}
	+ p[Q_{z^2}]
	\begin{bmatrix}
 	   & & & & &1\\
	   & & & &  &\\
 	  & & & & & \\
 	   & & & & & \\
 	    & & & & &
	\end{bmatrix}
	\nonumber\\
	\mathcal{B}^{\alpha \alpha,\alpha' \alpha'}_{ab,a'b'} &= \begin{aligned}[t]
	  & p_\textup{site}[Q_{x^2+y^2+z^2}]\,\delta_{ab}\delta_{a' b'}\\
	  & + p_\textup{site}[Q_{z^2}]\,\delta_{z,a} \delta_{z,b} \delta_{z, a'} \delta_{z, b'},
	\end{aligned}
\end{align}
where zero entries are omitted for simplicity.

When substituting
$\mathcal{B}^{\alpha \alpha,\alpha' \alpha'}$
back into the decision function, one finds that the first term in Eq.~\eqref{eq:SI-block_1111} just leads to the normalization of a spin $\mb{S}^{\alpha}$.
The second term, defining the product $(S^\alpha_z)^2 (S^{\alpha'}_z)^2$,
nevertheless relates to a nontrivial constraint when compared to the cross and bond-bond block in Fig.~\ref{fig:SI-PM}(c) and (d), which can be expressed as
\begin{align}
	\mathcal{B}^{\alpha \alpha,\alpha' \neq \beta'}_{a a^\prime} &= p_\textup{cross}[Q_{z^2}]\,\delta_{z,a} \delta_{z,b} \delta_{z, a'} \delta_{z, b'}, \\
	\mathcal{B}^{\alpha \neq \beta,\alpha' \neq \beta'}_{a a^\prime} &= p_\textup{bond}[Q_{z^2}]\,\delta_{z,a} \delta_{z,b} \delta_{z, a'} \delta_{z, b'}.
\end{align}
Note that the $p[Q_{x^2+y^2+z^2}]$ term does not appear here, because bond and cross terms do not obey an intrinsic constraint, unlike spin normalization in on-site terms.

\begin{table}
  \centering
  \begin{tabular*}{0.475\textwidth}{rcl@{\extracolsep{\fill}}*{5}{D{.}{.}{2.4}}}
  \toprule\addlinespace[-0.05em]
  \midrule
	& & & \multicolumn{1}{c}{$\gamma_{x^2+y^2+z^2}$} & \multicolumn{1}{c}{$\gamma_{z^2}$} & \multicolumn{1}{c}{$\gamma_{x^2+y^2}$} & \multicolumn{1}{c}{$\gamma_{x^2-y^2}$} & \multicolumn{1}{c}{$\gamma_{xy+yx}$}\\\midrule
	IV  &|& $T_\infty$ &        &        & -0.334 & -0.333 & -0.333\\
	III &|& $T_\infty$ &        &        & -0.331 &        &       \\
	III &|& IV         &        &        &        & -0.334 & -0.334\\
	V   &|& $T_\infty$ &        & -0.341 &        &        &       \\
	VI  &|& $T_\infty$ & -0.314 &        &        &        &       \\
  \midrule\addlinespace[-0.05em]
  \bottomrule
  \end{tabular*}
  \caption{Ratios of the relevant bond and on-site quadrupolar moments, $\gamma_\bullet\coloneqq \mean{Q^{\alpha\beta}_\bullet}/\mean{Q^{\alpha\alpha}_\bullet}$. These are calculated by taking the ratio between the weights of the relevant component with respect to ``bond'' and ``cross'' blocks, or ``cross'' and ``on-site'' blocks (see Eqs.~\eqref{eq:relation_zz}, \eqref{eq:relation_xy}, and \eqref{eq:relation_xyz}). Both ways yield consistent results in all cases (varying at most in the final digit). The weights of the ordering components themselves were calculated through a least-squares fit to the coefficient matrix as explained in Appendix~\ref{app:rank2_op}.}
  \label{tab:gamma}
\end{table}

We observe that these weights satisfy the relation,
\begin{align} \label{eq:relation_zz}
	 \gamma_{z^2}\coloneqq\frac{p_\textup{bond}[Q_{z^2}]}{p_\textup{cross}[Q_{z^2}]} = \frac{p_\textup{cross}[Q_{z^2}]}{p_\textup{site}[Q_{z^2}]} = \frac{\mean{S^\alpha_zS^\beta_z}}{\mean{S^\alpha_zS^\alpha_z}}
	 \approx -\frac{1}{3},
\end{align}
for $\alpha\neq\beta$ up to numerical accuracy (see Table \ref{tab:gamma}).
Summing over all sublattice indices ($\alpha,\beta$), $\gamma_{z^2}$ is absorbed by the ratio between on-site ($\alpha=\beta$; 4) and bond terms ($\alpha\neq\beta$; 12).
As a consequence, Eq.~\eqref{eq:relation_zz} in turn gives rise to the relation
\begin{align}
	\sum_{\alpha} \mean{S^\alpha_z S^\alpha_z}_\textup{cl}
	+ \sum_{\alpha \neq \beta} \mean{S^\alpha_z S^\beta_z}_\textup{cl}
	= \bigg\langle \kern-0.3em\bigg(\sum_{\alpha} S^{\alpha}_z\bigg)^{\kern-0.2em2} \bigg\rangle_\textup{\kern-0.3emcl} = 0,
\end{align} \label{eq:relation_Sz}
 where $\langle \dots \rangle_\textup{cl}$ averages over all tetrahedral clusters.

As $(\dots)^2$ is semi-positive definite, the constraint
\begin{align} \label{eq:cons_ice_rule}
	S^{(1)}_z + S^{(2)}_z + S^{(3)}_z + S^{(4)}_z = 0
\end{align}
has to be fulfilled for each tetrahedron individually.
Contrary to the spin normalization relation, which is an intrinsic constraint, this constraint emerges from the cooperative behavior of spins.
It defines the \emph{$2$-in-$2$-out} rule of spin ice.

Spin ice is known as an example of classical spin liquids.
It does not possess any long-range order, but features topological characteristics such as extensive ground state degeneracy (exGSD) and an effective $U(1)$ gauge-theoretical description~\cite{Henley10}.
These topological features actually are underpinned by the ice rule, Eq.~\eqref{eq:cons_ice_rule}.
Therefore, the TK-SVM is able to identify spin ice by means of the characteristic ice rule.

\subsubsection{Constraint in the easy-plane}
We continue the analysis of emergent constraints in other regions in the phase diagram Fig.~\ref{fig:fiedler}.
This involves the phases~III, IV, and VI.

In Sec.~\ref{sec:ops} we have discussed ordering components in the phases~III and IV.
We saw that phase~III has a planar quadrupolar component
$Q_{x^2 + y^2} = S^{\alpha}_x S^{\beta}_x + S^{\alpha}_y S^{\beta}_y$ which can be defined either by a single spin ($\alpha = \beta$) or on a bond connecting two spins ($\alpha \neq \beta$).

We find that the weights with which the site and bond terms manifest themselves also fulfill a relation
\begin{align} \label{eq:relation_xy}
	 \gamma_{x^2+y^2}\coloneqq\frac{p_\textup{bond}[Q_{x^2+y^2}]}{p_\textup{cross}[Q_{x^2+y^2}]} = \frac{p_\textup{cross}[Q_{x^2+y^2}]}{p_\textup{site}[Q_{x^2+y^2}]}
	 \approx -\frac{1}{3}.
\end{align}
(See Table~\ref{tab:gamma} and Appendix \ref{app:rank2_op} for details.)
Similar to the case of spin ice, the relation Eq.~\eqref{eq:relation_xy} in turn leads to a cooperation of spins
\begin{align}
\bigg\langle \bigg(\sum_{\alpha} S^{\alpha}_x\bigg)^2 \bigg\rangle_\textup{cl}
= \bigg\langle \bigg(\sum_{\alpha} S^{\alpha}_y\bigg)^2 \bigg\rangle_\textup{cl}
 = 0,
\end{align}
and, consequently, a vectorial constraint on the $S_x$ and $S_y$ components,
\begin{align} \label{eq:cons_xy}
\begin{pmatrix}
  S_x^{(1)} + S_x^{(2)} + S_x^{(3)} + S_x^{(4)}\\
  S_y^{(1)} + S_y^{(2)} + S_y^{(3)} + S_y^{(4)}
\end{pmatrix}
= \begin{pmatrix}
	0 \\
	0
\end{pmatrix}.
\end{align}
Therefore, phase~III  is not just an explicitly symmetry-breaking phase, but also subject to constrained dynamics.
This constraint is equivalent to that obtained by an irreducible-representation decomposition, Eq.~(5) in Ref.~\onlinecite{Taillefumier17}, from which one can derive pinch points in the spin structure factor.

The constraint in Eq.~\eqref{eq:relation_xy} is also observed in phase~IV (BSN phase), and is reflected by the weights of the biaxial orders,
$p[Q_{x^2-y^2}]$ and $p[Q_{xy+yx}]$ corresponding to the quadrupolar components defined in Eqs.~\eqref{eq:op_D2h_xy} and~\eqref{eq:op_C2h_xy}. We observe that both components occur with approximately the same weight for each type of block (see Appendix~\ref{app:rank2_op}), as well as ratios of $\gamma_{x^2-y^2}=\gamma_{xy+yx}=-1/3$ (see Table~\ref{tab:gamma}) among them.
Therefore, phase~IV is a constrained biaxial phase where a local order and an emergent constraint coexist.
As the latter signals an underlying gauge symmetry~\cite{Taillefumier17}, it also represents an instance where symmetry-breaking order coexists with an emergent gauge theory.

This coexistence also indicates a crucial difference between emergent spin nematic orders and intrinsic nematic orders in the context of liquid crystals.
In the latter case, nematic order parameters are considered as fundamental degrees of freedom (after coarse graining), whereas dipolar fields are typically trivial by construction.
In an emergent spin nematic phase on the other hand, even in the absence of a long-range dipolar order, ordinary spins may remain strongly constrained and exhibit nontrivial correlations.

\subsubsection{Isotropic constraint}

Lastly, in phase~VI, even though no local order is detected by the four-spin cluster up to rank $4$, we observe in $C_{\mu\nu} (\textup{VI}\,|\,T_{\infty})$ a relation between isotropic components,
\begin{align}
  Q^{\alpha\beta}_{x^2+y^2+z^2} &\coloneqq S^\alpha_xS^\beta_x +S^\alpha_yS^\beta_y + S^\alpha_zS^\beta_z
\end{align}
in the bond and cross terms,
\begin{align} \label{eq:relation_xyz}
	 \gamma_{x^2+y^2+z^2}\coloneqq\frac{p_\textup{bond}[Q_{x^2+y^2+z^2}]}{p_\textup{cross}[Q_{x^2+y^2+z^2}]} = -0.31,
\end{align}
whereas a similar ratio cannot be inferred from the on-site blocks where it is canceled by spin normalization.
However, contrary to Eq.~\eqref{eq:relation_zz} and Eq.~\eqref{eq:relation_xy}, here $\gamma_{x^2+y^2+z^2}$ is noticeably different from $-1/3$ and shows a dependence on training samples.
Such behavior seems to imply that the characterization of phase~VI is emerging but not yet sharply defined in the training samples.

\begin{figure}
  \centering
  \includegraphics{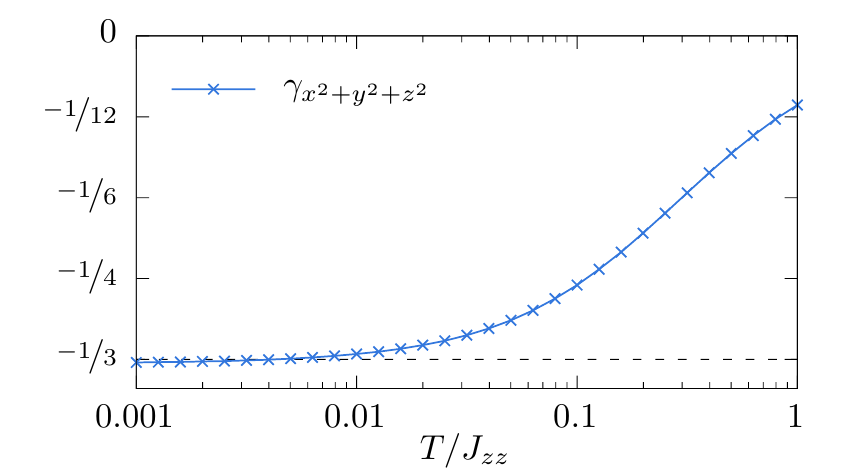}
  \caption{The ratio of the bond and on-site isotropic ordering components is plotted along the $J_\pm = -0.5J_{zz}$ line as a function of the temperature of the training samples.
  For $\gamma_{x^2+y^2+z^2}=-1/3$, the constraint $\mb{S}^{(1)}+\mb{S}^{(2)}+\mb{S}^{(3)}+\mb{S}^{(4)}=0$ for the four spins in each tetrahedron is perfectly fulfilled; for $\gamma_{x^2+y^2+z^2}=0$ the spins are entirely independent.
  Indeed, $\gamma_{x^2+y^2+z^2}$ is seen to approach $-1/3$ as $T\to 0$.
  Hence, the curve shows the crossover from a trivial paramagnet at high temperature to a cooperative paramagnet at low temperature.}
  \label{fig:gamma_phaf}
\end{figure}

To verify this conjecture and quantify the variation of $\gamma_{x^2+y^2+z^2}$,
we separately train the TK-SVM on each of $17$ parameter points along $J_{\pm} = -0.5J_{zz}$, which are apparently most representative of the phase, and fictitious isotropic configurations.
We extract $\gamma_{x^2+y^2+z^2}$ from the resultant coefficient matrices and plot it in Fig.~\ref{fig:gamma_phaf} against the temperature of the respective training data.
It becomes apparent that it approaches to $-1/3$ as $T\to 0$.

As a result, we can interpret the relation Eq.~\eqref{eq:relation_xyz} as a constraint that isotropically affects all of the three spin components,
\begin{align} \label{eq:cons_xyz}
\begin{pmatrix}
  S_x^{(1)} + S_x^{(2)} + S_x^{(3)} + S_x^{(4)} \\
  S_y^{(1)} + S_y^{(2)} + S_y^{(3)} + S_y^{(4)} \\
  S_z^{(1)} + S_z^{(2)} + S_z^{(3)} + S_z^{(4)}
\end{pmatrix}
= \begin{pmatrix}
	0 \\
	0 \\
	0
\end{pmatrix}.
\end{align}
However, this constraint is only obeyed in the ground state.
At finite temperature, it will be softened by thermal fluctuations and, consequently, a finite portion of spins will be released from the ground state configuration.

This is reminiscent of gapless excitations.
Indeed, at $J_{\pm} = -0.5J_{zz}$, the XXZ Hamiltonian Eq.~\eqref{eq:H_XXZ} becomes a pyrochlore Heisenberg model in local coordinates.
It is analog to the pyrochlore Heisenberg antiferromagnet (HAF) which is an example for gapless classical spin liquids (or cooperative paramagnets)~\cite{MoessnerChalker98a,MoessnerChalker98b}.
Moreover, the fluctuation-induced deviation of $\gamma$ from the ground state value $-1/3$ is also consistent with the finding that this phase has blurred pinch points at intermediate temperatures in the spin structure factor in Ref.~\onlinecite{Taillefumier17}.

Finally, the extracted order parameters and constraints confirm the hierarchy of the phases in Eq.~\eqref{eq:hierarchy} inferred from the bias criterion.
Phase~VI is ${\rm O}(3)$ symmetric, not breaking symmetry.
Nevertheless, owing to the constraint Eq.~\eqref{eq:cons_xyz}, it does not explore the entire configuration space, thus appears less disordered than the trivial paramagnet.
Phase~III and V feature constrained dynamics in easy-plane and easy-axis, respectively; both have the $D_{\infty h}$ symmetry.
Phase~IV breaks the symmetry of phase~III and develops a $C_{2h}$ coplanar order.
Furthermore, the bias criterion respects the distinct constraint in phase~IV and V, so does not assign them a rank, though the latter is more symmetric.
The magnetization $\mb{M}_{\perp}$ of phase~I has the $C_{1h}$ point-group symmetry and breaks the in-plane $\mathrm{O}(2)$ symmetry entirely.
Thereby, we can express the disorder hierarchy by the nature of the phases,
\begin{equation}\label{eq:hierarchy_final}
\begin{tikzcd}[column sep=tiny, row sep=tiny]
& & &
	\underset{\substack{\text{easy} \\ \text{plane}}}{D_{\infty h}} \rightarrow \underset{\text{biaxial}}{C_{2h}}
					\ar[dr, start anchor={[yshift=0.5em]}] & \\
\underset{\text{trival}}{\mathrm{O}(3)}\ar[rr] & &
		\underset{\text{constrained}}{\mathrm{O}(3)}
				\ar[ur, end anchor={[yshift=1em]}] \ar[dr] & &
					\underset{\text{magnetic}}{C_{1h}}.
						\\
& & &
		\underset{\substack{\text{easy}\\\text{axis}}}{D_{\infty h}}
			\ar[ur]
\end{tikzcd}
\end{equation}

\section{Thermodynamics of constraints}\label{sec:d_functions}

When learning the phase diagram, Fig.~\ref{fig:fiedler}, we pointed out that gradual change of Fiedler values at phase boundaries implies a crossover.
Thereby, aside from the topology of the phase diagram, the graph analysis also provides an intuitive way of recognizing crossovers and the regions where they take place.
In this section, we will confirm this interpretation by examining the analytical order parameters and constraints extracted from the coefficient matrices.
Moreover, we will also discuss possible advantages of those quantities in the identification of phase transitions and crossovers, as compared to the use of conventional quantities such as heat capacity or magnetic susceptibility.

\subsection{Single crossover}

We first discuss the crossover between the spin ice (phase~V) and the trivial paramagnet (phase~II).
This crossover is well understood in terms of the Schottky anomaly in the specific heat which may serve as a reference point for the quantity learned by TK-SVM.

The analytical quantity extracted from $C_{\mu\nu}(\textup{V}\,|\,T_\infty)$,
Eqs.~\eqref{eq:SI-block_1111}\--\eqref{eq:relation_zz},
may be expressed as
\begin{align}\label{eq:Gamma_zz}
	\Gamma_{z}
	&\coloneqq \bigg\langle
	\frac{1}{4} \sum_\alpha (S^\alpha_z)^2
	- \frac{1}{12} \sum_{\alpha \neq \beta} S^\alpha_z S^\beta_z
	-\frac{1}{3}
	\bigg\rangle_{\rm cl} \\
	&= \bigg\langle
	\frac{1}{3} \sum_\alpha (S^\alpha_z)^2
	- \frac{1}{12} \Bigl(\sum_{\alpha} S^\alpha_z\Bigr)^2
	-\frac{1}{3}
	\bigg\rangle_{\rm cl}.
\end{align}
It measures the fulfillment of the ice-rule Eq.~\eqref{eq:cons_ice_rule} and may be regarded as an order parameter, where
$\Gamma_{z}  = 1$ if the ice-rule is fully satisfied,
while $\Gamma_{z}  = 0$ for uncorrelated spins.
$\Gamma_z$ is normalized to satisfy these limiting cases.

We can define a susceptibility to quantify the fluctuation of $\Gamma_{z}$,
\begin{align}
	\chi[\Gamma_{z}] \coloneqq \frac{1}{T} \big(
	\langle \Gamma_{z}^2 \rangle - \langle \Gamma_{z} \rangle^2 \big).
\end{align}
One expects that $\chi[\Gamma_{z}]$ is smooth at crossovers, but exhibits discontinuity or divergence when experiencing phase transitions.

\begin{figure}
  \centering
  \includegraphics[scale=0.95]{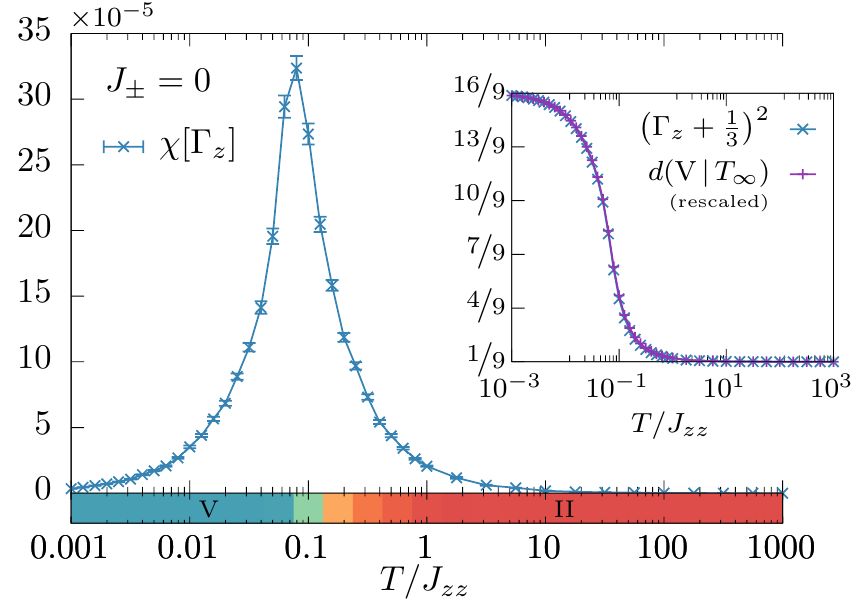}
  \caption{The susceptibility of the spin ice constraint $\Gamma_z$ [Eq.~\eqref{eq:Gamma_zz}] is measured along the $J_{zz}=0$ line. The colored bar below the abscissa shows the corresponding slice of the phase diagram, Fig.~\ref{fig:fiedler}, for comparison. The peak of the susceptibility is seen to coincide with the locus of the crossover between the spin ice phase (blue) and the paramagnet (red). The inset confirms the agreement of the decision function with $(\Gamma_z+1/3)^2$, the expression that was inferred from its coefficient matrix, given suitable affine rescaling of the former.}
  \label{fig:Gamma_zz}
\end{figure}

In Fig.~\ref{fig:Gamma_zz}, we measure $\Gamma_{z}$ and $\chi[\Gamma_{z}]$ along the $J_{\pm} = 0$ line.
We indeed observe that $(\Gamma_z+1/3)^2$ collapses onto the decision function, verifying our interpretation of the coefficient matrix $C_{\mu\nu}(\textup{V}\,|\,T_\infty)$.
[Note that the constant $-1/3$ in the definition \eqref{eq:Gamma_zz} is not included in the TK-SVM decision function,~\emph{i.e.} $d(\textup{V}\,|\,T_\infty)\propto (\Gamma_z+1/3)^2+\textup{const.}$]
Moreover, $\chi[\Gamma_{z}]$ shows a broad peak at the boundary between the two phases,
indicating a crossover driven by thermal violation of the ice rule.
As expected, the characteristic temperature of this peak agrees with that inferred from the Schottky anomaly (the dashed line between phases~II and V in Fig.~\ref{fig:fiedler})~\cite{Taillefumier17}.

The above example confirms our approach. Next, we apply it to the crossover between the cooperative (VI) and the trivial paramagnet, whose characterization is less clear.
The (normalized) order parameter corresponding to the relation~\eqref{eq:relation_xyz} and the isotropic constraint Eq.~\eqref{eq:cons_xyz} is given by
\begin{align}
	\Gamma_{xyz} \coloneqq 3 \sum_a {\Gamma_a} = 1 - \frac{1}{4}\Big\langle\Big\|\sum_{\alpha}\mb{S}^\alpha\Big\|^2\Big\rangle_\textup{cl},\label{eq:Gamma_xyz}
\end{align}
where
\begin{align}
\Gamma_a \coloneqq \bigg\langle
	\frac{1}{3} \sum_\alpha (S^\alpha_a)^2
	- \frac{1}{12} \Bigl(\sum_{\alpha} S^\alpha_a\Bigr)^2
	-\frac{1}{3}
	\bigg\rangle_{\rm cl},
\end{align}
and $a = x, y, z$.
Accordingly, we define its susceptibility
$	\chi[\Gamma_{xyz}] = \big(
	\langle \Gamma_{xyz}^2 \rangle - \langle \Gamma_{xyz} \rangle^2 \big)/T$.

\begin{figure}
  \centering
  \includegraphics[scale=0.95]{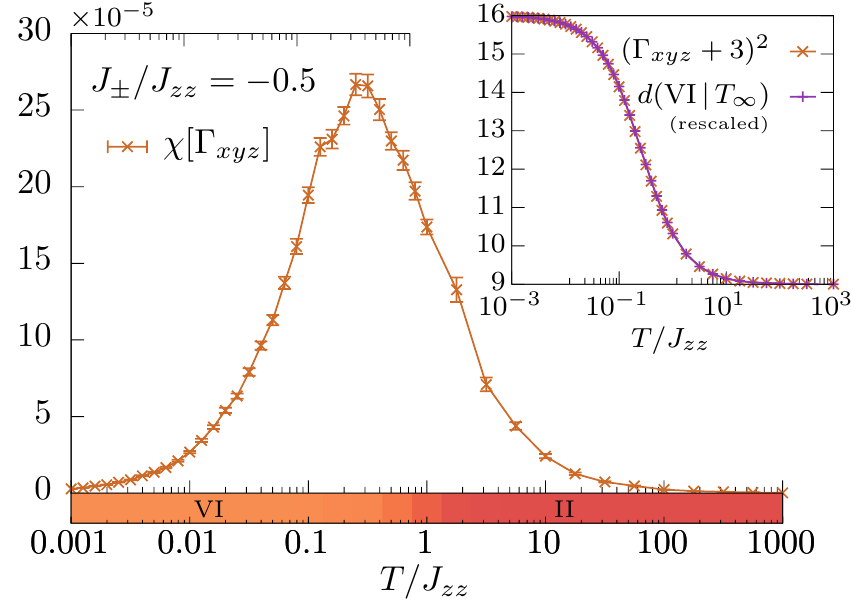}
  \caption{The susceptibility associated with the isotropic constraint $\Gamma_{xyz}$ [Eq.~\eqref{eq:Gamma_xyz}] is measured along the $J_{zz}=-0.5J_{zz}$ line (Heisenberg limit). The colored bar below the abscissa shows the corresponding slice of the phase diagram, Fig.~\ref{fig:fiedler}, for comparison. The broad peak of the susceptibility is seen to coincide with the locus of the crossover between the cooperative (orange) and the trivial (red) paramagnet. The inset confirms the agreement of the decision function with $(\Gamma_{xyz}+3)^2$, the expression that was inferred from its coefficient matrix, given suitable affine rescaling of the former.}
  \label{fig:Gamma_xyz}
\end{figure}

$\Gamma_{xyz}$ and $\chi[\Gamma_{xyz}]$ are measured along the $J_{\pm} = -0.5J_{zz}$,~\emph{i.e.} in the Heisenberg limit, in Fig.~\ref{fig:Gamma_xyz}.
One observes a bump over approximately three orders of magnitude in temperature.
The location of its maximum is in agreement with that inferred on the basis of the magnetic susceptibility~\cite{Taillefumier17},
its profile nevertheless marks a much larger area influenced by the crossover.
We note that this region has in fact been hinted at by the slow variance of the Fiedler vector entries between the two phases.

In addition, the behavior of $\chi[\Gamma_{xyz}]$ is also consistent with that of the ratio $\gamma$ in Fig.~\ref{fig:gamma_phaf}.
Hence, this spin liquid is only well defined in the regime where $\gamma \approx -1/3$ at $T \lesssim 0.01$.
Thereafter, the crossover starts to take hold, until very high temperature.

Here, we have relied on the susceptibilities of the analytical order parameters which were previously extracted from the coefficient matrices learned by TK-SVM.
This has the advantage that they are immediately relatable to physical quantities.
One may also choose to rely on the decision function directly and define a susceptibility for it instead.
The resulting quantity cannot immediately be converted into the susceptibilities of the order parameter, whereas the reverse is true (the analytical order parameter can be squared and compared to the decision function as is the case in the insets of Figs.~\ref{fig:Gamma_zz} and \ref{fig:Gamma_xyz}).
Nonetheless, the susceptibilities of the decision functions exhibit peaks at approximately the same positions.
We point out that this approach is entirely feasible to get a first impression of the behavior, without the effort of dissecting the coefficient matrices first.

\subsection{Sequence of phase transitions and crossovers}
The quantities learned by the TK-SVM are optimized to distinguish two given phases.
This specialization can lead to a higher sensitivity of identifying phase transitions and crossovers, in particular when the system involves multiple fluctuating fields.

This is exemplified by the phase transition and crossovers relating to phases~III and IV.
In Sec.~\ref{sec:patterns}, we saw that phase~IV is characterized by the quadrupolar fields $Q^\mathrm{U}_{C_{2h}}$ and $\mathbf{Q}^\mathrm{B}_{C_{2h}}$,
where the latter represents the symmetry-breaking order,
and the constraint Eq.~\eqref{eq:cons_xy} on the $S_x$ and $S_y$ components.
When entering phase~III, the biaxial order parameter $\mathbf{Q}^\mathrm{B}_{C_{2h}}$ vanishes, but the constraint remains in place.
This constraint can be defined by an order parameter
\begin{align}
	\Gamma_{xy} \coloneqq \frac{3}{2}(\Gamma_{x} + \Gamma_{y}),\label{eq:Gamma_xy}
\end{align}
which is distinct from that of the cooperative paramagnet ($\Gamma_{xyz}$) at higher temperature.

Therefore, by increasing temperature at $J_{\pm} < -0.5J_{zz}$ in the phase diagram  Fig.~\ref{fig:fiedler},
the system undergoes the upper branch of the sequence in Eq.~\eqref{eq:hierarchy_final} (excluding the $C_{1h}$ phase).
We expect the order parameter fields
$\mathbf{Q}^\mathrm{B}_{C_{2h}}, \Gamma_{xy}$, and $\Gamma_{xyz}$ to respond to the corresponding phase transition and crossovers separately.

\begin{figure}
  \centering
  \includegraphics{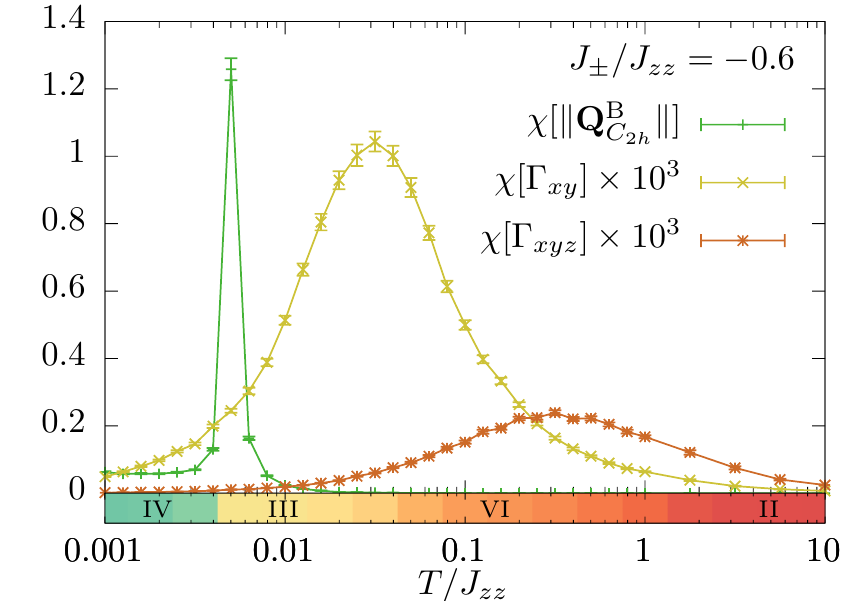}
  \caption{Various susceptibilities are measured along the $J_{zz}=-0.6J_{zz}$ line. The colored bar below the abscissa shows the corresponding slice of the phase diagram, Fig.~\ref{fig:fiedler}, for comparison. The sharp peak in susceptibility of the biaxial order parameter $\mathbf{Q}^\mathrm{B}_{C_{2h}}$ [Eq.~\eqref{eq:C2h_op}] pinpoints the phase transition between the generalized biaxial (green) and uniaxial (yellow) phases; the broad peaks in the susceptibilities of the order parameters derived from the constraints, $\Gamma_{xy}$ [Eq.~\eqref{eq:Gamma_xy}] and $\Gamma_{xyz}$ [Eq.~\eqref{eq:Gamma_xyz}], locate the crossovers into the cooperative (orange) and trivial (red) paramagnet, respectively. Note that the latter are exaggerated by a factor of 1000.}
  \label{fig:Gamma_xy}
\end{figure}

In Fig.~\ref{fig:Gamma_xy}, their corresponding susceptibilities,
$\chi[\|\mathbf{Q}^\mathrm{B}_{C_{2h}}\|]$,
$\chi[\Gamma_{xy}]$,
and $\chi[\Gamma_{xyz}]$, are measured along the $J_\pm = -0.6J_{zz}$ line.
Indeed, they exhibit individual peaks/bumps at the relevant transitions and crossovers.
The pronounced peak in $\chi[\|\mathbf{Q}^\mathrm{B}_{C_{2h}}\|]$ identifies the generalized biaxial-uniaxial phase transition~\cite{Liu17} between the constrained $C_{2h}$ and $D_{\infty h}$ phase.
The bumps in $\chi[\Gamma_{xy}]$ and $\chi[\Gamma_{xyz}]$
are responsible for the two subsequent crossovers.

This ability to isolate phase transitions and crossovers is to be contrasted with the analysis of conventional thermodynamic quantities.
For instance, the specific heat encodes thermal fluctuations of all order parameters at the same time; hence, not every phase transition or crossover may manifest itself noticeably.
In particular, signals of crossovers are potentially drowned out by a phase transition or may not be distinguishable from other nearby crossovers.
On the other hand, the magnetic susceptibility is sensitive to dipolar orders, nonetheless, it may not respond to the fluctuations of multipolar fields.

\section{Summary and outlook}\label{sec:summary}

Frustrated spin and spin-orbital-coupled systems accommodate a wide array of exotic phases, whereas their identification is usually difficult.
In this work, we demonstrated that support vector machines equipped with a tensorial kernel (TK-SVM) and combined with spectral graph partitioning can automatically learn the intricate phase diagram of a classical frustrated magnet.
While the method was originally proposed for detecting symmetry-breaking orders and phase transitions~\cite{Greitemann19, Liu19}, here we showed that it is also capable of detecting the emergent local constraints characteristic of spin liquids, and crossovers between different disordered states.

The method was applied to the XXZ model on the pyrochlore lattice.
This model hosts several unconventional phases, including a hidden spin nematic order and three different classical spin liquids~\cite{Taillefumier17,Benton18}, and thus poses a challenging test case for the verification of our method.

We systematically illustrated the utility of the SVM bias parameter in the graph analysis to construct the phase diagram of the XXZ model (Sec.~\ref{sec:graph}).
The correct topology of the phase diagram (Fig.~\ref{fig:fiedler}) was obtained without prior information on either order parameters or constraints.
This represents an extremely valuable feature of our method as such information is often unavailable or less obvious in frustrated systems.
In addition to the topology, the graph analysis also provides a way to visualize extended regions in which crossovers take place across up to two orders of magnitude in temperature.

We also elucidated in detail how to interpret the machine results for the analytical characterization of the phases.
By virtue of the strong interpretability of the kernel, we were able to extract order parameters (Sec.~\ref{sec:ops}) and emergent constraints (Sec.~\ref{sec:constraints}) from the SVM coefficient matrices, thus identifying the nature of the respective phases.
For each local constraint, a susceptibility was introduced to measure the associated crossover (Sec.~\ref{sec:d_functions}).
Moreover, a hierarchy of phases was also derived to systematically discuss their relative levels of disorder [Eq.~\eqref{eq:hierarchy_final}].

The successful application to the XXZ pyrochlore antiferromagnet indicates that TK-SVM can serve as an efficient framework for constructing intricate phase diagrams.
It could be particularly useful for systems involving multiple competing interactions.

One class of such systems are the Kitaev materials~\cite{Janssen19, Takagi19,Trebst17}, including spin liquid candidates $\mathrm{Li_2 Ir O_3}$, $\mathrm{Na_2 Ir O_3}$~\cite{Chaloupka10, Nussinov15}, $\mathrm{Sr_2 Ir O_4}$~\cite{Jackeli09} and $\alpha\text{-}\mathrm{RuCl_3}$~\cite{Banerjee16, Kim16}.
These materials involve various competing interactions, such as the Heisenberg, Kitaev, and off-diagonal Gamma interactions and also show a strong dependence on external magnetic fields~\cite{Zheng17, Wolter17, Janssen19}.
Even when varying only a subset of the interaction parameters, their (classical) phase diagrams exhibit multiple exotic orders competing with the desired spin liquids~\cite{Janssen17, Chern19}, whereas order parameters of those phases are not yet clear and significant parts of the phase diagram remain unexplored.
Following the framework provided by TK-SVM to compute the phase diagram, one does not have to scan each physical parameter individually; phase transitions and crossovers in the entire parameter space of interest can instead be identified in one fell swoop.
Independent of the number of physical parameters, the analysis will result in a univariate histogram (see Fig.~\ref{fig:fiedler_histo}), whose peaks imply distinct phases.
One can thereby obtain a comprehensive phase diagram more efficiently.

The ability to construct a high-dimensional phase diagram in turn allows to systematically classify which interactions favor hidden order or spin liquids.
This may accelerate the pace at which researchers can scrutinize material-inspired model Hamiltonians and might open the door towards the engineering of unconventional phases in the future.

\begin{acknowledgements}
JG, KL, and LP acknowledge support from FP7/ERC Consolidator Grant QSIMCORR, No. 771891, and the Deutsche Forschungsgemeinschaft (DFG, German Research Foundation) under Germany's Excellence Strategy -- EXC-2111 -- 390814868.
LJ acknowledges hospitality from LMU Munich and was supported by the Agence Nationale de la Recherche under Grant No. ANR-18-CE30-0011-01.
HY and NS acknowledge support from the Theory of Quantum Matter Unit of the Okinawa Institute of Science and Technology Graduate University.
HY is further supported by the Japan Society for the Promotion of Science (JSPS) Research Fellowships for Young Scientists.
Our simulations make use of the $\nu$-SVM formulation~\cite{Scholkopf00}, the LIBSVM library~\cite{Chang01, Chang11}, and the ALPS\-Core library~\cite{Gaenko17}.
\end{acknowledgements}

\begin{appendix}
\section{Details of the Monte Carlo simulation}
\label{app:latt}

There are four sublattices on the pyrochlore lattice, whose positions in a given tetrahedron are
\begin{align}
\begin{aligned}
\mathbf{r}_0 &= \dfrac{a}{8} \left(+1, +1,+1 \right),
&
\mathbf{r}_1 &= \dfrac{a}{8} \left(+1,-1,-1 \right),\\
\mathbf{r}_2 &= \dfrac{a}{8} \left(-1,+1,-1 \right),
&
\mathbf{r}_3 &= \dfrac{a}{8} \left(-1,-1,+1 \right),
\end{aligned}
\label{eq:r}
\end{align}
where $a$ is the length of the traditional cubic unit cell made of 16 sites. The local coordinate frame of the spin components $\mathbf{S}_i = (S_{i,x}, S_{i,y}, S_{i,z})$ is defined on each sublattice as follows
\begin{align}
\begin{aligned}
\mathbf{x}_0 &= \dfrac{1}{\sqrt{6}}(-2,+1,+1),
&
\mathbf{x}_1 &= \dfrac{1}{\sqrt{6}}(-2,-1,-1),\\
\mathbf{x}_2 &= \dfrac{1}{\sqrt{6}}(+2,+1,-1),
&
\mathbf{x}_3 &= \dfrac{1}{\sqrt{6}}(+2,-1,+1),
\end{aligned}
\label{eq:local-easy-plane-x}
\end{align}
\begin{align}
\begin{aligned}
\mathbf{y}_0 &= \dfrac{1}{\sqrt{2}}(0,-1,+1),
&
\mathbf{y}_1 &= \dfrac{1}{\sqrt{2}}(0,+1,-1),\\
\mathbf{y}_2 &= \dfrac{1}{\sqrt{2}}(0,-1,-1),
&
\mathbf{y}_3 &= \dfrac{1}{\sqrt{2}}(0,+1,+1),
\end{aligned}
\label{eq:local-easy-plane-y}
\end{align}
and
\begin{align}
\begin{aligned}
\mathbf{z}_0 &= \dfrac{1}{\sqrt{3}}(+1,+1,+1),
&
\mathbf{z}_1 &= \dfrac{1}{\sqrt{3}}(+1,-1,-1),\\
\mathbf{z}_2 &= \dfrac{1}{\sqrt{3}}(-1,+1,-1),
&
\mathbf{z}_3 &= \dfrac{1}{\sqrt{3}}(-1,-1,+1).
\end{aligned}
\label{eq:local-111-axis}
\end{align}

A heat-bath algorithm for single-spin-flip updates was combined with overrelaxation and parallel tempering. Preliminary thermalization is carried out in two steps: first a slow annealing from high temperature to the temperature of measurement $T$ during $t_{e}$ Monte Carlo steps (MCs) followed by $t_{e}$ MCs at temperature $T$. After thermalization, measurements are done every 10 MCs during $t_{m}=10 t_{e}$ MCs. Typical Monte Carlo times range from $t_{m}=10^{6}$ to $3\times 10^{7}$ Monte Carlo sweeps.

\section{Graph weighting}
\label{app:graph_weights}

In our original proposal of the graph analysis using TK-SVM~\cite{Liu19}, a threshold value $\rho_c$ was employed to decide whether a given edge should be included in the graph if its corresponding bias $\rho$ fulfills $||\rho|-1| > \rho_c$.
In the same paper, we made the observation that it is generally favorable to choose large values of $\rho_c$, thereby applying a more restrictive criterion resulting in a sparsely populated graph. Indeed, we proposed to tune $\rho_c$ to the largest possible value for which the resulting graph is still connected.

While this approach works well enough when the phase diagram is composed of a few symmetry-breaking phases, it does discard information on the strength of the connection and different transitions or crossovers may become apparent at different values of $\rho_c$. Rather than using a fixed cutoff to construct an unweighted graph, in this work we have opted to use the bias value of each edge to determine its weight, resulting in a weighted graph.

\begin{figure}
  \centering
  \includegraphics{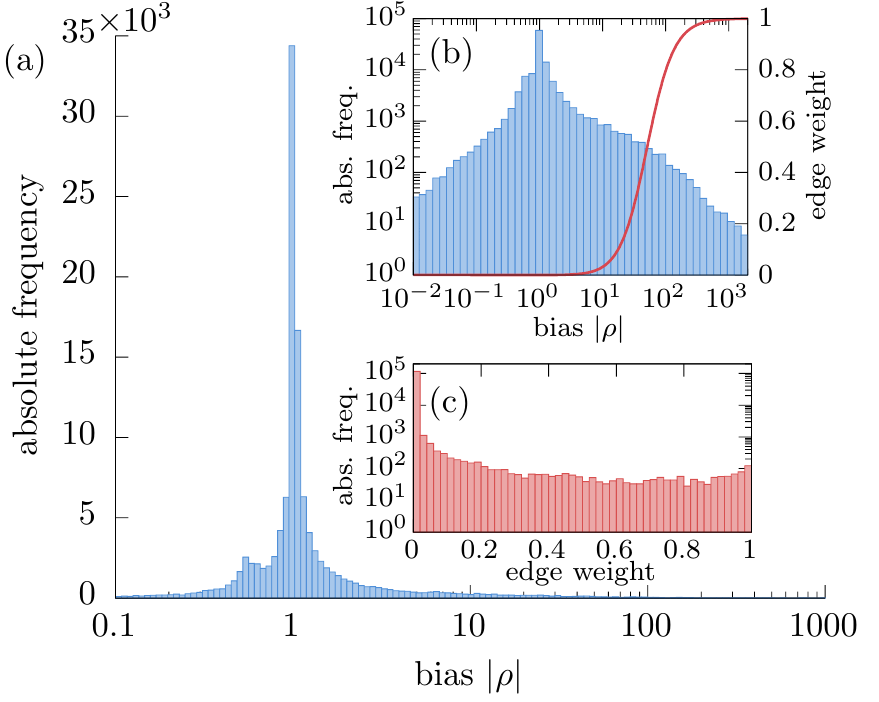}
  \caption{(a, b) Histogram of the (unoriented) biases obtained from TK-SVM among the 493 points in parameter space. The bin width is proportional to the bias values and their absolute frequency is plotted on a (a) linear and (b) logarithmic scale, respectively. The red curve in panel (b) is the Lorentzian weighting function, Eq.~\eqref{eq:lorentzian} with characteristic scale $\rho_c=50$, used to map biases to graph edge weights. Panel (c) shows an (again logarithmic) histogram of the resulting edge weight distribution.}
  \label{fig:bias_histo}
\end{figure}

The biases of all the potential edges span many orders of magnitude as can be seen from their histogram in Fig.~\ref{fig:bias_histo}(a). Aside from the pronounced peak at $|\rho|=1$ corresponding to the edges spanning across phases, the remaining biases are following a fat-tailed distribution which becomes apparent from the log-log version of the same histogram in the inset Fig.~\ref{fig:bias_histo}(b).

\begin{figure*}
  \centering
  \includegraphics[width=0.88\textwidth]{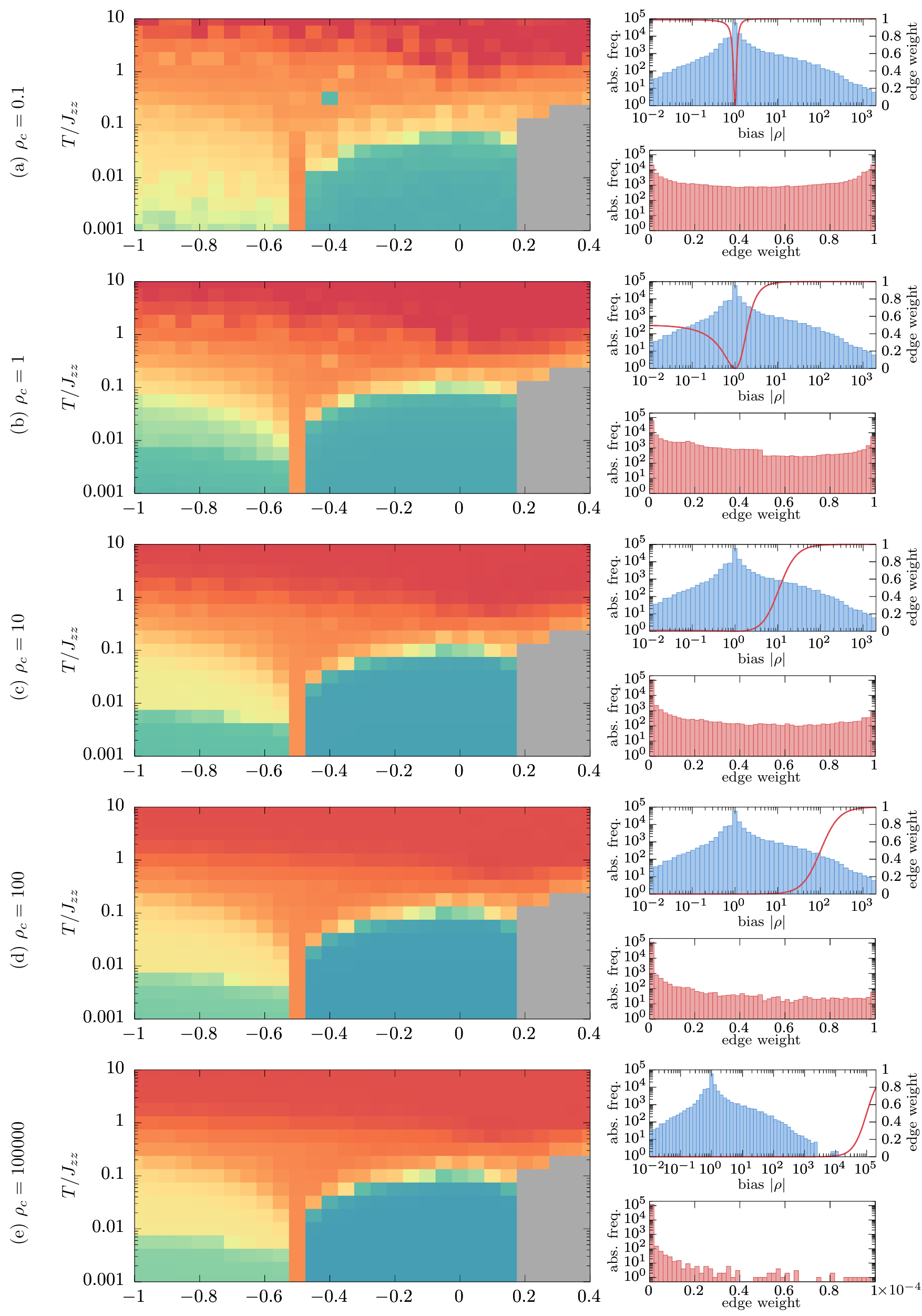}
  \caption{The Fiedler vector, representing the extracted phase diagram, is shown along with histograms of the biases and corresponding edge weights for a variety of different choices of the characteristic scale $\rho_c$ of the weighting function. The weighting functions are superimposed on the bias histograms in each case. The color scale used for the Fiedler vector is identical to that used in Fig.~\ref{fig:fiedler}.}
  \label{fig:rhoc_sweep}
\end{figure*}

Using the biases (or its deviation from unity) directly as weights for the edges does not yield useful results, as it overrepresents the importance of large biases.
Instead, we employ a Lorentzian weighting function,
\begin{align}
  w(\rho) &= 1 - \frac{\rho_c^2}{(|\rho|-1)^2+\rho_c^2},\label{eq:lorentzian}
\end{align}
to map biases to the interval $[0,1)$. $\rho_c$ again gives a characteristic scale up to which edges do not contribute significantly. The resulting distribution of edge weights covers the unit interval somewhat uniformly.
We note that our previous approach amounted to choosing a weighting function $\Theta(||\rho|-1|-\rho_c)$ instead, resulting in weights zero and one exclusively.
Choosing a continuous weighting function instead retains more of the information provided by the biases and also simplifies the spectral analysis by ensuring that the graph never exhibits any truly disconnected components as these would lead to a degenerate eigenvalue zero of the Laplacian matrix.
This is of particular practical relevance in identifying the crossovers (where vertices have few and weak connections to their neighbors) correctly as such; using a hard threshold, these vertices are at risk of being completely disconnected from the surrounding phases.

The Fiedler vector obtained from the graph is robust with respect to the choice of the weighting function and its characteristic scale $\rho_c$.
Fig.~\ref{fig:rhoc_sweep} shows the Fiedler vector along with the distribution of the biases and their corresponding weights for different values of $\rho_c$ spanning six orders of magnitude.
Even in the most inclusive approach in panel (a), the topology of the phase diagram can be recognized, even though the result is generally more noisy and the phase boundary of the spin nematic phase appears blurred. As successively more edges are discarded, these shortcomings are gradually rectified. The resulting Fiedler vectors for $\rho_c=10^2$ [panel (d)], $10^3$, $10^4$, and $10^5$ [panel (e)] are virtually indistinguishable, even though in the latter case, only the largest of biases contribute to the graph appreciably. Beyond $\rho_c=10^6$, the analysis is limited by the numerical accuracy of the weights. Considering that this last choice exceeds the maximum bias that was found, it is unnatural.

We conclude that (at least in the present case) the choice of the weighting function is not of practical concern. We have reaffirmed our previous statement that it is generally beneficial to be more exclusive when deciding which biases to include in the graph. Using a continuous weighting function is, however, crucial to ensure that the graph stays (weakly) connected.

\section{Pooling training data}
\label{app:pooling}

Having inferred the phase diagram from the analysis of the graph, we proceed by relabeling the training data according to the phase labels~I--VI.

To that end, we have identified the corresponding intervals of the entries of the Fiedler vector and indicate these in Fig.~\ref{fig:fiedler_histo}. Their choice was guided by the by the location of the peaks in the histogram in the case of regions~II--V. Since region~VI does not exhibit a pronounced peak for the reasons discussed in Sec.~\ref{sec:graph}, we centered the interval on the value attained in the Heisenberg limit ($J_\pm=-J_{zz}/2$, $T\to 0$).

\begin{figure}
  \centering
  \includegraphics{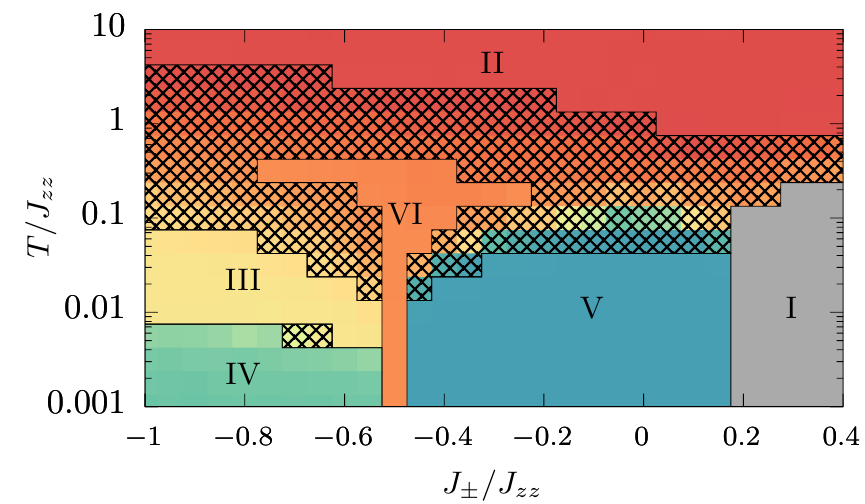}
  \caption{The partitioning of the phase diagram is superimposed on the Fiedler vector (see Fig.~\ref{fig:fiedler}). Regions which can be unambiguously assigned to one of the phases, labeled by roman numerals~I--VI, are separated by solid lines. For the hatched regions, this is not possible unambiguously and the corresponding data are excluded from the ensuing analysis.}
  \label{fig:pooling}
\end{figure}

The entries which do not fall into either of these intervals can be attributed to crossovers in between. The corresponding training data cannot be labeled unambiguously and are therefore not included in the pooled data set. The final partitioning of the phase diagram is presented in Fig.~\ref{fig:pooling}. [We point out that we have deviated slightly from the strict interval-based partitioning laid out in this appendix by manually excluding a handful of points on the boundary of region~V (which are intermediate crossover points, but might otherwise have been labeled III or IV, purely based on their Fiedler vector entry).]

After pooling, the SVM is trained with $24,\!000$ (I), $49,\!500$ (II), $16,\!500$ (III), $18,\!000$ (IV), $43,\!500$ (V), and $13,\!500$ (VI) spin configurations, respectively, while $81,\!500$ samples have been excluded.
In addition, $50,\!000$ fictitious random ($T_\infty$) spin configurations are included as the control group.

\section{Extracting quadrupolar moments}\label{app:rank2_op}

\begin{table*}
  \centering
  \begin{tabular*}{\textwidth}{@{\extracolsep{\fill}}cc*{5}{D{.}{.}{2.4}}}
    \toprule\addlinespace[-0.05em]
    \midrule
    & & \multicolumn{1}{c}{$p[Q_{x^2+y^2+z^2}]$} & \multicolumn{1}{c}{$p[Q_{z^2}]$} & \multicolumn{1}{c}{$p[Q_{x^2+y^2}]$} & \multicolumn{1}{c}{$p[Q_{x^2-y^2}]$} & \multicolumn{1}{c}{$p[Q_{xy+yx}]$}\\\midrule
                    & on-site & -0.564 & -0.021 & \multicolumn{1}{B{.}{.}{2.4}}{ 1.274} & \multicolumn{1}{B{.}{.}{2.4}}{ 0.278} & \multicolumn{1}{B{.}{.}{2.4}}{ 0.280}\\
    IV | $T_\infty$ & cross   & -0.007 &  0.006 & \multicolumn{1}{B{.}{.}{2.4}}{-0.425} & \multicolumn{1}{B{.}{.}{2.4}}{-0.093} & \multicolumn{1}{B{.}{.}{2.4}}{-0.093}\\
                    & bond    &  0.003 & -0.002 & \multicolumn{1}{B{.}{.}{2.4}}{ 0.142} & \multicolumn{1}{B{.}{.}{2.4}}{ 0.031} & \multicolumn{1}{B{.}{.}{2.4}}{ 0.031}\\\midrule
                    & on-site &  0.717 &  0.281 & \multicolumn{1}{B{.}{.}{2.4}}{-1.681} & -0.001 & -0.001\\
    III | $T_\infty$& cross   &  0.097 & -0.081 & \multicolumn{1}{B{.}{.}{2.4}}{ 0.555} &  0.000 &  0.000\\
                    & bond    & -0.028 &  0.024 & \multicolumn{1}{B{.}{.}{2.4}}{-0.184} &  0.000 &  0.000\\\midrule
                    & on-site & -0.088 &  0.082 &  0.146 & \multicolumn{1}{B{.}{.}{2.4}}{ 0.933} & \multicolumn{1}{B{.}{.}{2.4}}{ 0.941}\\
    III | IV        & cross   &  0.022 & -0.021 & -0.052 & \multicolumn{1}{B{.}{.}{2.4}}{-0.311} & \multicolumn{1}{B{.}{.}{2.4}}{-0.314}\\
                    & bond    & -0.005 &  0.005 &  0.019 & \multicolumn{1}{B{.}{.}{2.4}}{ 0.104} & \multicolumn{1}{B{.}{.}{2.4}}{ 0.105}\\\midrule
                    & on-site & -0.086 & \multicolumn{1}{B{.}{.}{2.4}}{ 1.068} & -0.073 &  0.000 &  0.000\\
    V | $T_\infty$  & cross   & -0.018 & \multicolumn{1}{B{.}{.}{2.4}}{-0.364} &  0.016 &  0.000 &  0.000\\
                    & bond    &  0.003 & \multicolumn{1}{B{.}{.}{2.4}}{ 0.124} & -0.003 &  0.000 &  0.000\\\midrule
                    & on-site &  0.001 & -0.001 &  0.000 &  0.000 &  0.000\\
    VI | $T_\infty$ & cross   & \multicolumn{1}{B{.}{.}{2.4}}{ 0.997} &  0.000 &  0.000 &  0.000 &  0.000\\
                    & bond    & \multicolumn{1}{B{.}{.}{2.4}}{-0.313} &  0.000 &  0.000 &  0.000 &  0.000\\
    \midrule\addlinespace[-0.05em]
    \bottomrule
  \end{tabular*}
  \caption{The weights of the tentative quadrupolar ordering components, $p[Q_\bullet]$, are tabulated for five classifiers which are analyzed in the main text. These weights were obtained through a least-squares fit based on all blocks in the full coefficient matrix of each of the site-site (``on-site''), site-bond (``cross''), and bond-bond (``bond'') types, as discussed in Sec.~\ref{sec:constraints}.
  For each classifier, weights corresponding to components which contribute significantly (and are of physical relevance) are set in bold type. The ratios between these weights for the three block types are given in Table~\ref{tab:gamma}.}
  \label{tab:components}
\end{table*}


From the general structure of the blocks $\mathcal{B}$ in the coefficient matrix, Eq.~\eqref{eq:block_C2h}, parameterized by variables $A$ through $E$, we identify five constituent patterns:
{\allowdisplaybreaks
\begin{align}
  \mathcal{A}[Q_{x^2+y^2+z^2}] &= \begin{bmatrix}
     1  & \zz & \zz & \zz &  1  & \zz & \zz & \zz &  1 \\
    \zz & \zz & \zz & \zz & \zz & \zz & \zz & \zz & \zz\\
    \zz & \zz & \zz & \zz & \zz & \zz & \zz & \zz & \zz\\
    \zz & \zz & \zz & \zz & \zz & \zz & \zz & \zz & \zz\\
     1  & \zz & \zz & \zz &  1  & \zz & \zz & \zz &  1 \\
    \zz & \zz & \zz & \zz & \zz & \zz & \zz & \zz & \zz\\
    \zz & \zz & \zz & \zz & \zz & \zz & \zz & \zz & \zz\\
    \zz & \zz & \zz & \zz & \zz & \zz & \zz & \zz & \zz\\
     1  & \zz & \zz & \zz &  1  & \zz & \zz & \zz &  1
  \end{bmatrix},\label{eq:p1}\\
  \mathcal{A}[Q_{z^2}] &= \begin{bmatrix}
    \zz & \zz & \zz & \zz & \zz & \zz & \zz & \zz &  1 \\
    \zz & \zz & \zz & \zz & \zz & \zz & \zz & \zz & \zz\\
    \zz & \zz & \zz & \zz & \zz & \zz & \zz & \zz & \zz\\
    \zz & \zz & \zz & \zz & \zz & \zz & \zz & \zz & \zz\\
    \zz & \zz & \zz & \zz & \zz & \zz & \zz & \zz & \zz\\
    \zz & \zz & \zz & \zz & \zz & \zz & \zz & \zz & \zz\\
    \zz & \zz & \zz & \zz & \zz & \zz & \zz & \zz & \zz\\
    \zz & \zz & \zz & \zz & \zz & \zz & \zz & \zz & \zz\\
    \zz & \zz & \zz & \zz & \zz & \zz & \zz & \zz & \zz
  \end{bmatrix},\\
  \mathcal{A}[Q_{x^2+y^2}] &= \begin{bmatrix}
    \zz & \zz & \zz & \zz & \zz & \zz & \zz & \zz & \zz\\
    \zz & \zz & \zz & \zz & \zz & \zz & \zz & \zz & \zz\\
    \zz & \zz & \zz & \zz & \zz & \zz & \zz & \zz & \zz\\
    \zz & \zz & \zz & \zz & \zz & \zz & \zz & \zz & \zz\\
     1  & \zz & \zz & \zz &  1  & \zz & \zz & \zz & \zz\\
    \zz & \zz & \zz & \zz & \zz & \zz & \zz & \zz & \zz\\
    \zz & \zz & \zz & \zz & \zz & \zz & \zz & \zz & \zz\\
    \zz & \zz & \zz & \zz & \zz & \zz & \zz & \zz & \zz\\
     1  & \zz & \zz & \zz &  1  & \zz & \zz & \zz & \zz
  \end{bmatrix},\\
  \mathcal{A}[Q_{x^2-y^2}] &= \begin{bmatrix}
    \zz & \zz & \zz & \zz & \zz & \zz & \zz & \zz & \zz\\
    \zz & \zz & \zz & \zz & \zz & \zz & \zz & \zz & \zz\\
    \zz & \zz & \zz & \zz & \zz & \zz & \zz & \zz & \zz\\
    \zz & \zz & \zz & \zz & \zz & \zz & \zz & \zz & \zz\\
    -1  & \zz & \zz & \zz &  1  & \zz & \zz & \zz & \zz\\
    \zz & \zz & \zz & \zz & \zz & \zz & \zz & \zz & \zz\\
    \zz & \zz & \zz & \zz & \zz & \zz & \zz & \zz & \zz\\
    \zz & \zz & \zz & \zz & \zz & \zz & \zz & \zz & \zz\\
     1  & \zz & \zz & \zz & -1  & \zz & \zz & \zz & \zz
  \end{bmatrix},\\
  \mathcal{A}[Q_{xy+yx}] &= \begin{bmatrix}
    \zz & \zz & \zz & \zz & \zz & \zz & \zz & \zz & \zz\\
    \zz & \zz & \zz & \zz & \zz & \zz & \zz & \zz & \zz\\
    \zz & \zz & \zz & \zz & \zz & \zz & \zz & \zz & \zz\\
    \zz & \zz & \zz & \zz & \zz & \zz & \zz & \zz & \zz\\
    \zz & \zz & \zz & \zz & \zz & \zz & \zz & \zz & \zz\\
    \zz &  1  & \zz &  1  & \zz & \zz & \zz & \zz & \zz\\
    \zz & \zz & \zz & \zz & \zz & \zz & \zz & \zz & \zz\\
    \zz &  1  & \zz &  1  & \zz & \zz & \zz & \zz & \zz\\
    \zz & \zz & \zz & \zz & \zz & \zz & \zz & \zz & \zz
  \end{bmatrix}.
\end{align}}

Each of these patterns contributes a quadrupolar ordering component $Q_\bullet$ to the decision function, which is related to the corresponding pattern $\mathcal{A}[Q_\bullet]$ through the relation $\big(Q_\bullet^{\alpha\beta}\big)^2 = \Tr\left[\mathcal{A}[Q_\bullet]\,(\mb{S}^\alpha\otimes\mb{S}^\beta)^{\otimes 2}\right]$:
\begin{align}
  Q^{\alpha\beta}_{x^2+y^2+z^2} &= S^\alpha_xS^\beta_x + S^\alpha_yS^\beta_y + S^\alpha_zS^\beta_z,\\
  Q^{\alpha\beta}_{z^2} &= S^\alpha_zS^\beta_z,\\
  Q^{\alpha\beta}_{x^2+y^2} &= S^\alpha_x S^\beta_x + S^\alpha_y S^\beta_y,\\
  Q^{\alpha\beta}_{x^2-y^2} &=  S^\alpha_x S^\beta_x - S^\alpha_y S^\beta_y,\\
  Q^{\alpha\beta}_{xy+yx} &= S^\alpha_x S^\beta_y + S^\alpha_y S^\beta_x,
\end{align}
where the first line corresponds to the intrinsic normalization constraint, $\|\mb S^\alpha\|^2=1$ in case of ``on-site'' ($\alpha=\beta$) blocks.

The weights $p[Q_\bullet]$ of the ordering components are thus given by decomposing $\mathcal{B}$ in terms of $\mathcal{A}[Q_\bullet]$,~\emph{i.e.} by solving the linear equations $\bds{\mathcal{A}}\bds{p} = \mathcal{B}$, where
$\bds{\mathcal{A}} = (\mathcal{A}[Q_{x^2+y^2+z^2}], \dots, \mathcal{A}[Q_{xy+yx}])$ and $\bds{p} = (p[Q_{x^2+y^2+z^2}], \dots, p[Q_{xy+yx}])^T$.
In case the form of Eq.~\eqref{eq:block_C2h} was followed exactly, this would result in five independent equations.
When $\mathcal{B}$ is rather obtained through the SVM coefficient matrix and, hence, noisy, the linear system is overdetermined and the optimal choice of the component weights can be found by a least-squares fit,~\emph{i.e.} by minimizing $\|\bds{\mathcal{A}}\bds{p}-\mathcal{B}\|^2$.
In fact, since blocks of each type (``on-site'', ``cross'', and ``bond'' type) occur in the full coefficient matrix $C_{\mu\nu}$ many times over, all of these instances can be included in the least-squares fit to obtain common weights for each of the three block types.

The results are tabulated in Table~\ref{tab:components} for the various cases discussed in the main text (Figs.~\ref{fig:SN}, \ref{fig:SL}, \ref{fig:SI-PM}).
In each case, the ordering components which contribute significantly support the previous characterization of phases~IV, III, and V.
One also notes that in the situations involving the biaxial phase~IV, the ordering components $Q_{x^2-y^2}$ and $Q_{xy+yx}$ occur with equal weight (up to numerical precision), in line with the fact that they form the biaxial order parameter, $\mb{Q}^\mathrm{B}_{C_{2h}}$.
For each significant component, the ratio between ``on-site'' and ``cross'' blocks and ``cross'' and ``bond'' blocks, respectively, is approximately equal and given in Table~\ref{tab:gamma}.

As expected, the pattern~\eqref{eq:p1} contributes significantly only in on-site blocks where it corresponds to a constant and is, thus, physically irrelevant. The exception is phase~VI where it is the only pattern to occur and does so in cross and bond blocks. Hence, it does contribute nontrivially to the decision function in a way such that $d\propto (\Gamma_{xyz}-\gamma^{-1})^2+\textup{const.}$, where $\gamma = p_\textup{cross}[Q_{x^2+y^2+z^2}] / p_\textup{bond}[Q_{x^2+y^2+z^2}]$ and $\Gamma_{xyz}$ is the isotropic constraint, Eq.~\eqref{eq:Gamma_xyz}.

\end{appendix}

\bibliographystyle{apsrev4-1}
\bibliography{svm-pyro}
\end{document}